\documentclass[11pt]{article}

\newif\ifsubmission

\usepackage{fontawesome5}

\usepackage[preprint]{acl}

\usepackage{times}
\usepackage{latexsym}

\usepackage[T1]{fontenc}

\usepackage[utf8]{inputenc}

\usepackage{microtype}
\usepackage{amssymb}

\usepackage{inconsolata}

\usepackage{graphicx}

\usepackage{booktabs} 
\usepackage{tabularx} 
\usepackage{multirow} 
\usepackage{xspace}
\usepackage{adjustbox} 
\usepackage{amsmath} 
\usepackage{xcolor}
\usepackage{setspace}
\usepackage{colortbl} 
\usepackage{caption}
\usepackage{subcaption} 
\usepackage{float}
\usepackage[most]{tcolorbox}
\usepackage{listings}
\usepackage{capt-of} 
\usepackage{ragged2e}

\tcbuselibrary{skins, breakable, listings, theorems}
\usepackage{enumitem}
\newcommand{\ie}{{\emph{i.e.}}}
\newcommand{\eg}{{\emph{e.g.}}}

\newcommand{\vs}{\emph{vs.}}

\newcommand{\name}{\textsc{OS-Symphony}}

\definecolor{lightgray}{gray}{0.9} 
\definecolor{darkgray}{gray}{0.5} 
\definecolor{lightred}{RGB}{255, 120, 120}   
\definecolor{lightgreen}{RGB}{100, 200, 100} 
\definecolor{lightblue}{RGB}{100, 150, 255}  
\definecolor{myblue}{RGB}{219,232,247}
\definecolor{lakeblue}{RGB}{0, 128, 192}
\definecolor{lightyellow}{rgb}{1.0, 1.0, 0.88}  
\definecolor{lightpink}{rgb}{1.0, 0.71, 0.76}   
\definecolor{lightorange}{rgb}{1.0, 0.83, 0.62} 
\definecolor{lightcyan}{rgb}{0.88, 1.0, 1.0}   
\definecolor{lightpurple}{rgb}{0.9, 0.8, 1.0}   
\definecolor{lavender}{rgb}{0.9, 0.8, 1.0}       
\definecolor{lilac}{rgb}{0.78, 0.64, 0.8}        
\definecolor{periwinkle}{rgb}{0.8, 0.8, 1.0}     
\definecolor{mauve}{rgb}{0.87, 0.63, 0.87}       
\definecolor{orchid}{rgb}{0.85, 0.44, 0.84}      
\definecolor{amethyst}{rgb}{0.6, 0.4, 0.8}       
\definecolor{wisteria}{rgb}{0.79, 0.63, 0.86}    
\definecolor{dustylavender}{rgb}{0.76, 0.7, 0.86} 
\definecolor{frenchlavender}{rgb}{0.8, 0.6, 0.7} 
\definecolor{heliotrope}{rgb}{0.87, 0.73, 1.0}   
\definecolor{plum}{rgb}{0.87, 0.63, 0.87}        
\definecolor{LakeBlue}{RGB}{0,61,153}

\definecolor{veronica-red}{RGB}{196,30,58}

\DeclareCaptionType{prompt}[Prompt][Prompt]
\lstdefinestyle{longpromptstyle}{
    basicstyle=\ttfamily 
    breaklines=true,
    columns=flexible,
    frame=none,
    language=[LaTeX]TeX,
    aboveskip=0pt,
    belowskip=0pt
}

\title{
    \name: A Holistic Framework for Robust and Generalist Computer-Using Agent
}

\author{
\textbf{Bowen Yang}\textsuperscript{$1,2$}\thanks{\, Equal Contribution.},
\textbf{Kaiming Jin}\textsuperscript{$3\,*$},
\textbf{Zhenyu Wu}\textsuperscript{$2$},
\textbf{Zhaoyang Liu}\textsuperscript{$4$},
\textbf{Qiushi Sun}\textsuperscript{$5$}, 
\textbf{Zehao Li}\textsuperscript{$2$}, \\
\textbf{Jingjing Xie}\textsuperscript{$6$}, 
\textbf{Zhoumianze Liu}\textsuperscript{$2$},
\textbf{Fangzhi Xu}\textsuperscript{$7$},
\textbf{Kanzhi Cheng}\textsuperscript{$8$},
\textbf{Qingyun Li}\textsuperscript{$9$}, 
\textbf{Yian Wang}\textsuperscript{$3$}, \\
\textbf{Yu Qiao}\textsuperscript{$2$},
\textbf{Zun Wang}\textsuperscript{$2$},
\textbf{Zichen Ding}\textsuperscript{$2$}\thanks{\, Corresponding Author.} \\
\textsuperscript{$1$}University of Science and Technology of China
\textsuperscript{$2$}Shanghai AI Laboratory \\
\textsuperscript{$3$}National University of Singapore 
\textsuperscript{$4$}The Hong Kong University of Science and Technology \\
\textsuperscript{$5$}The University of Hong Kong
\textsuperscript{$6$}CUHK MMLab 
\textsuperscript{$7$}Xi'an Jiaotong University \\
\textsuperscript{$8$}Nanjing University 
\textsuperscript{$9$}Harbin Institute of Technology 
}

\begin{document}
\maketitle

\begin{abstract}

While Vision-Language Models~(VLMs) have significantly advanced Computer-Using Agents (CUAs), current frameworks struggle with robustness in long-horizon workflows and generalization in novel domains. These limitations stem from a lack of granular control over historical visual context curation and the absence of visual-aware tutorial retrieval.
To bridge these gaps, we introduce \name, a holistic framework that comprises an Orchestrator coordinating two key innovations for robust automation:
1) a Reflection-Memory Agent that utilizes milestone-driven long-term memory to enable trajectory-level self-correction, effectively mitigating visual context loss in long-horizon tasks; 2) Versatile Tool Agents featuring a Multimodal Searcher that adopts a \textit{SeeAct} paradigm to navigate a browser-based sandbox to synthesize live, visually aligned tutorials, thereby resolving fidelity issues in unseen scenarios.
Experimental results demonstrate that \name\ delivers substantial performance gains across varying model scales, establishing new state-of-the-art results on three online benchmarks, notably achieving 65.84\% on OSWorld.
\ifsubmission
    All research assets will be made publicly available.
\else
    Our code and project are publicly available at \href{https://github.com/OS-Copilot/OS-Symphony}{\faGithub\ OS-Copilot/OS-Symphony} and \href{https://os-copilot.github.io/OS-Symphony/}{\faGlobe\ OS-Symphony Homepage}.
\fi
\end{abstract}

\section{Introduction}

The landscape of digital task automation has been reshaped by the advancement of Vision-Language Models (VLMs)~\citep{bai2025qwen25vl,bai2025qwen3,wang2025internvl35,anthropic2025claude4.5,hong2025glm}, 
leading to vision-guided Computer-Using Agents (CUAs)~\citep{sun2024osgenesis, qin2025uitars, wang2025uiTARS2, xie2025scalingcomputerusegroundinguser, wu2025oracle, liu2025scalecua,wu2025gui}. 
By leveraging visual perception to interact with digital environments, these agents have expanded the scope and applicability of general-purpose automation.

While native CUAs~\citep{zhang2024large,hu2025agents,liu2025scalecua,wang2025opencuaopenfoundationscomputeruse} trained on large-scale computer-using trajectories are capable of digital navigation tasks, they often struggle to generalize to complex scenarios under a single-agent paradigm. Consequently, modular CUA frameworks~\citep{wu2024copilot,song2025coact,gonzalez2025unreasonable,guo2025agentic} have emerged by orchestrating multiple specialized sub-agents, \eg, planner, grounder, and coder~\citep{chen2025map,jia2025agentstore}, to coordinate seamlessly, exhibiting significant potential for developing reliable generalist CUAs.
\begin{figure}[t]
    \centering
    \includegraphics[width=\columnwidth]{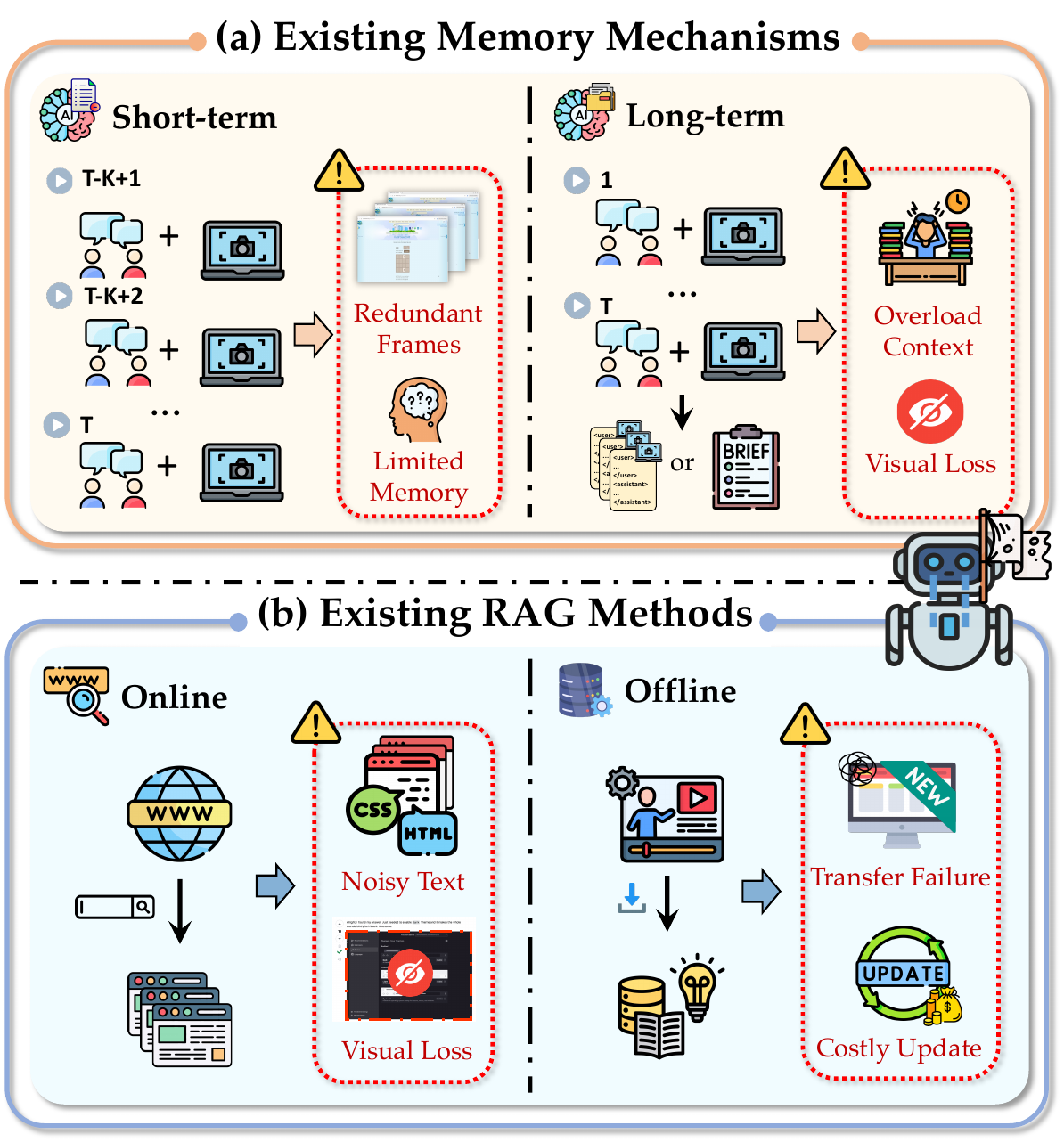} 
    \captionsetup{skip=0pt, position=bottom}
    \caption{Current limitations in CUA framework.}
    \label{fig:teaser}
    \vspace{-0.5cm}
\end{figure}

Despite the promising progress in agentic frameworks, they face two critical challenges. 
First, while memory modules~\cite{song2025coact, cheng2025mga, tian2025agentprog} are employed to support long-horizon tasks, current context management mechanisms often lack granular control over historical visual information curation and pruning. As illustrated in Fig.~\ref{fig:teaser}(a), this deficiency results in suboptimal utilization of historical visual information, rendering agents ill-equipped to identify potential errors like intent drift or cyclic behaviors. This lack of retrospective insight ultimately prevents the generation of meaningful reflections to refine planning in complex, long-horizon tasks.
Second, several works~\cite{xu2024agenttrek,sun2024multi,agent-s, guo2025agentic, xu2025retrieval} incorporate external knowledge via Retrieval-Augmented Generation (RAG) in an effort to generalize to unseen scenarios; however, as shown in Fig.~\ref{fig:teaser}(b), they either excessively rely on unimodal information, thereby overlooking vital semantic cues in the visual modality, or depend on local knowledge bases that incur high maintenance costs and struggle to adapt to new software. Consequently, these approaches fail to achieve robust generalization on out-of-distribution (OOD) tasks.

To this end, we propose \name, a holistic CUA framework comprising a decision-making \textbf{Orchestrator} coordinating two core designs to bridge these gaps:
1) \textbf{Reflection-Memory Agent} that leverages long-term memory to retain key ``milestone'' screenshots alongside abstract trajectories, effectively mitigating visual context loss. By visually auditing historical states, the RMA generates critical trajectory-level reflections according to a structured message protocol, providing high-level guidance for the Orchestrator to ensure robust performance over long-horizon tasks.
2) \textbf{Versatile Tool Agents}, highlighted by a meticulously designed \textit{Multimodal Searcher} alongside a Coder and Grounders that work synergistically to execute complex tasks. Specifically, our Searcher enables acquiring diverse tutorials via browser-based sandbox autonomously. By integrating visual information with spatial layouts, it provides high-fidelity, relevant tutorials, enabling the Orchestrator to leverage external multimodal knowledge for OOD scenarios.

We demonstrate the effectiveness of \name\ across diverse scales and benchmarks, achieving substantial performance leaps over current state-of-the-art methods with scores of 65.8\% on OSWorld ($\uparrow$2.4\%), 63.5\% on WindowsAgentArena ($\uparrow$6.9\%), and 46.0\% on MacOSArena ($\uparrow$38.0\%).
Beyond quantitative results, our rigorous ablation and granular analysis dissect the core drivers of this performance, offering valuable directions for future CUA development.

Our contributions are summarized as follows:

1) We propose \name, a holistic CUA framework which investigates a robust and generalist paradigm via collaboration of diverse agents to solve complicated tasks in practice.

2) We design a Reflection-Memory Agent to address the lack of granular control over historical visual context curation. By integrating milestone-driven long-term memory with a structured auditing protocol, it generates in-depth reflection for robust long-horizon planning.

3) We develop a suite of tool agents which facilitate solving tasks effectively. To overcome the absence of visual-aware tutorial retrieval, among these tools, \textit{Multimodal Searcher} is highlighted to harvest rich multimodal knowledge for OOD tasks by actively navigating the web pages.

4) Extensive evaluations across diverse operating systems and model scales validate the superior performance of \name. Furthermore, our framework empowers open-source VLMs to successfully execute long-horizon or unseen tasks that previously challenged their capabilities.
\section{Related Work}
\begin{figure*}[htbp]
    \centering
    \ifsubmission
    \includegraphics[width=\textwidth]{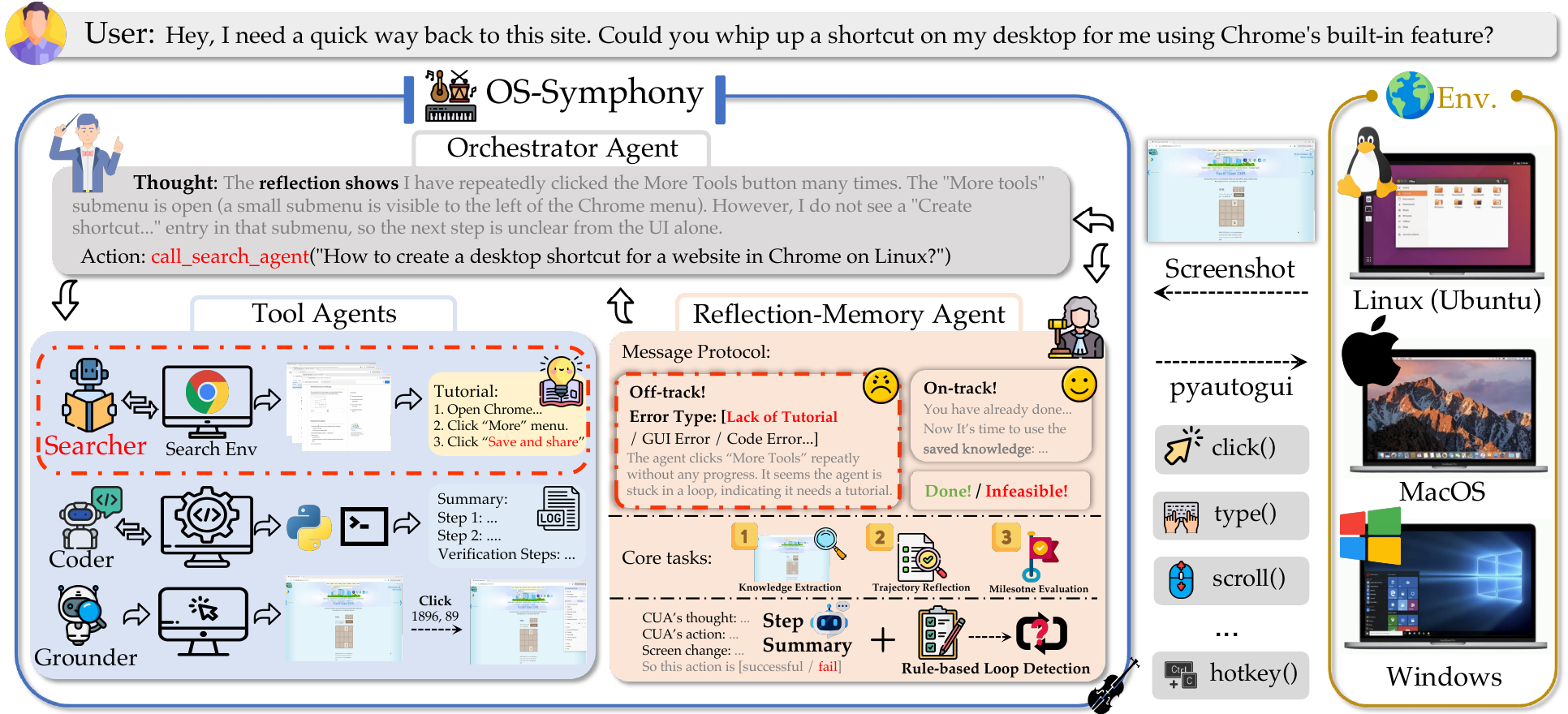} 
    \else
    \includegraphics[width=\textwidth]{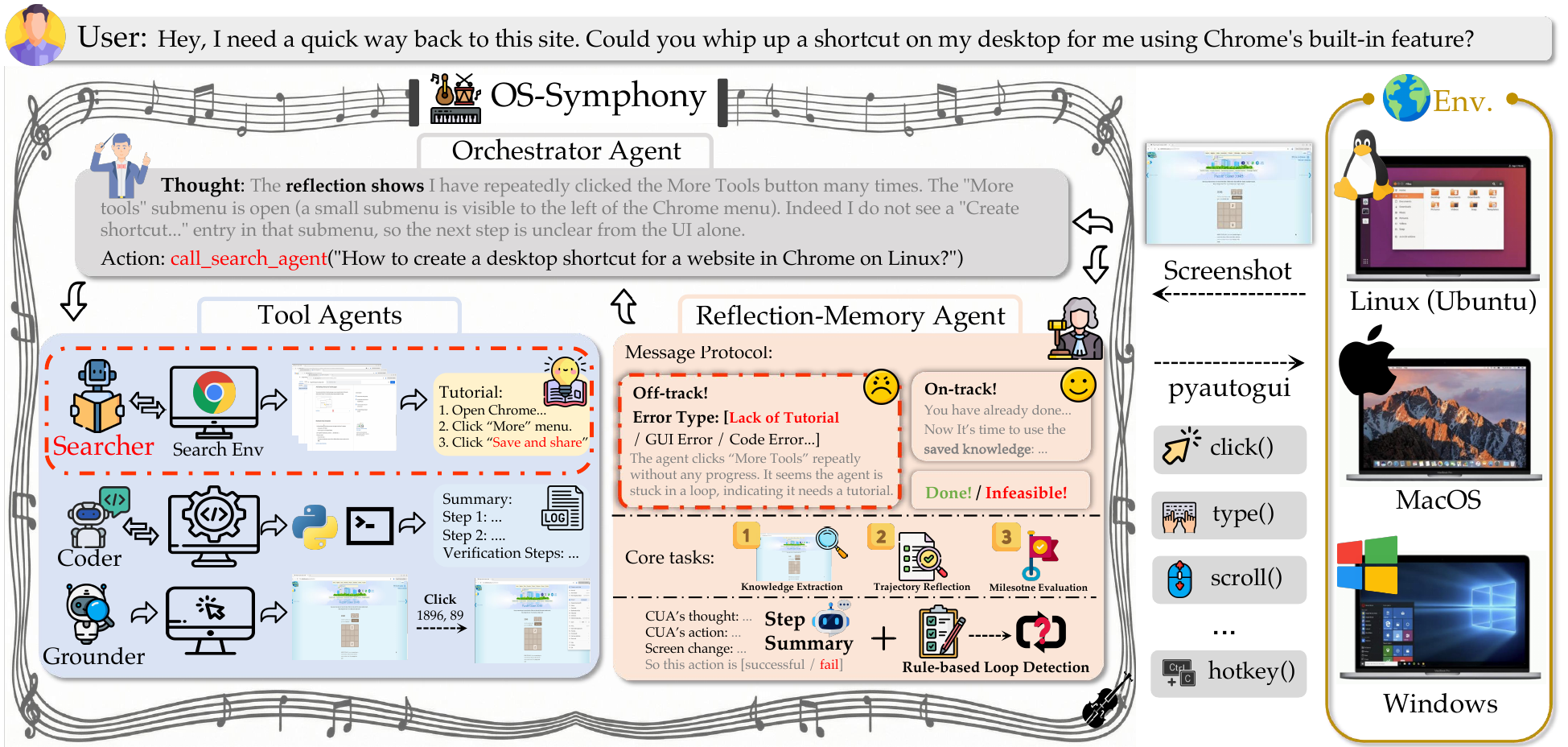} 
    \fi
    \vspace{-2mm}
    \captionsetup{skip=0pt, position=bottom}
    \caption{\textbf{Pipeline overview.} \name\ comprises three primary components: (1) The \textbf{Orchestrator}, acting as the system's brain, responsible for task understanding and action prediction; (2) Tool Agents, 
    consisting of \textbf{Grounder}, \textbf{Coder}, and \textbf{Searcher}, where the Searcher retrieves up-to-date tutorials in a human-like manner; and (3) The \textbf{Reflection-Memory Agent},
    which compresses execution trajectories to maintain long-term memory and facilitate trajectory-level reflection.}
    
    \label{fig:method_overview}
    \vspace{-0.5cm}
\end{figure*}
\noindent \textbf{Computer-Using Agents (CUAs).}
With the rapid development of Vision-Language Models~(VLMs) ~\citep{anthropic2025claude4, comanici2025gemini, openai2025o3systemcard, wang2025internvl35, bai2025qwen3}, Computer-Using Agents have become a novel paradigm to explore Human-Computer Interaction. 
Native CUAs pursue end-to-end digital autonomy,
encompassing both general-purpose models~\citep{openai2025computerusing, guo2025seed1, anthropic2025claude4.5} adapted for agentic tasks and specialized models~\citep{cheng2024seeclick,wu2024osatlas,xu2024aguvis,qin2025uitars,wang2025opencuaopenfoundationscomputeruse} fine-tuned on large-scale GUI datasets for dedicated computer use.
In parallel, CUA frameworks~\citep{wu2024copilot,Agent-S2,yang2025gta1, wu2025backtrackagent,ye2025mobile,zhang2025guimid} prioritize modularity by decomposing complex tasks. 
This approach enhances capability through modular collaboration while mitigating the data dependency of end-to-end training.
Beyond architectural paradigms, the field is transitioning from purely vision-based approaches~\citep{zhang2025api,zhang2025ufo2,wang2025uiTARS2} toward hybrid GUI-API strategies. While some methods~\citep{sun2024code,song2025coact,gonzalez2025unreasonable} leverage general-purpose interfaces like code execution, others~\citep{lai2025computerrl, yang2025ultracua, jia2025osworldmcp} depend on software-specific APIs via protocols such as MCP. However, relying on specific APIs often hinders generalization to niche or proprietary software. Consequently, retrieval-augmented strategies have emerged as a critical solution to bridge this gap.

\noindent \textbf{RAG for CUAs.}
To strengthen the generalization of CUAs, an increasing number of studies~\citep{zhang2024large, nguyen2025gui, hu2025agents} has integrated Retrieval-Augmented Generation~(RAG) to access external knowledge. Early lines of research~\citep{shi2025gui, agent-s, mei2025r, guo2025agentic} rely on general-purpose AI search engines~(\eg, Perplexica\footnote{\url{https://github.com/ItzCrazyKns/Perplexica}}) to perform static knowledge retrieval prior to task execution. In parallel, other efforts construct task-specific knowledge databases or training corpora by manually curating software documentation~\citep{xu2024agenttrek, xu2025retrieval} and video tutorials~\citep{zhao2025worldgui}. More recently, inspired by the DeepResearch paradigm~\citep{gemini2025deep,team2025tongyi,grok2025deeper,tao2025webshaper,wu2025webdancer, wu2025webwalker}, emerging methods have begun to tightly integrate GUI interaction with deep research capabilities, enabling agents to mine web-scale resources for complex reasoning~\citep{wang2025uiTARS2,wang2025gametar}.
However, purely text-based retrieval is insufficient for GUI scenarios, as it struggles to interpret verbose HTML and screenshot-heavy tutorials~\citep{li2025nestedbrowseruselearningagentic}. This highlights an urgent need to incorporate visual contexts into RAG tailored for GUI agents.

\noindent \textbf{Memory for CUAs.}
Efficient memory management is pivotal for long-horizon tasks~\citep{hu2025memoryageaiagents, agent_memory_survey_2024, hu2025agents,wang2024gui,sun2025scienceboard}. Recent LLM-based methods leverage history summarization~\citep{wu2025resum, yu2025memagent} or context folding~\citep{ye2025agentfold, sun2025scaling} to compress interactions. 
Other approaches leverage multi-scale memory~\citep{wu2024copilot, li2025deepagent}, fusing short-term working memory and long-term procedural memory to maintain a comprehensive repository of trajectories and acquired knowledge.
In GUI domain, some approaches~\citep{cheng2025mga,tian2025agentprog,sun2025coda} further distill trajectories, including screenshots, thoughts, and actions, into structured summaries. However, these approaches remain largely text-centric, often losing the visual semantics that are essential for progress tracking and downstream decision-making.
\section{\name}
\label{sec:method}

 In this section, we present the overall framework of \name, which comprises three synergistic components: an Orchestrator, a Reflection-Memory Agent~(RMA), and specialized Tool Agents. In the case shown in Fig.~\ref{fig:method_overview}, the Orchestrator first interprets feedback from the RMA, which identifies execution stagnation caused by a Chrome version mismatch between the task environment and the VLM's pre-training knowledge, together with a \textit{Lack of Tutorial} error. It then invokes the Searcher to retrieve a relevant tutorial, and following this guidance, successfully completes the task, achieving a closed-loop self-improvement cycle.

\subsection{Orchestrator}

We employ an Orchestrator, which serves as the core component, responsible for task interpretation, coordinating all Tool Agents, and finally selects an action. Formally, we model the decision-making process of the Orchestrator as:
\begin{equation}
    t_{i}, a_{i} = \mathcal{F}_{O}(\mathcal{I}, \mathcal{R}_{i}, o_{i}, \mathcal{T}, \mathcal{H}_{\text{short}}),
\end{equation}
where $\mathcal{F}_O$ represents the Orchestrator and $t_{i}$ and $a_{i}$ denote the thought and action components of the agent's output. Additionally, $\mathcal{I}$ represents the instruction, $\mathcal{R}_{i}$ signifies the reflection feedback provided by the RMA, $o_{i}$ denotes the current screenshot, and $\mathcal{T}$ is the retrieved tutorial, which is provided upon invoking the Searcher and is empty otherwise. Notably, $\mathcal{H}_{\text{short}}=\{(o_j, t_j, a_j)\}_{j=i-K+1}^{i-1}$ denotes the short-term memory. Assuming the significance of historical interactions decays with time, we restrict the Orchestrator's memory to a sliding window of the last $K$ turns, capturing the immediate dialogue and screenshots essential for precise next-action prediction.

\subsection{Reflection-Memory Agent}

\begin{figure*}[t]
    \centering
    \includegraphics[width=\textwidth]{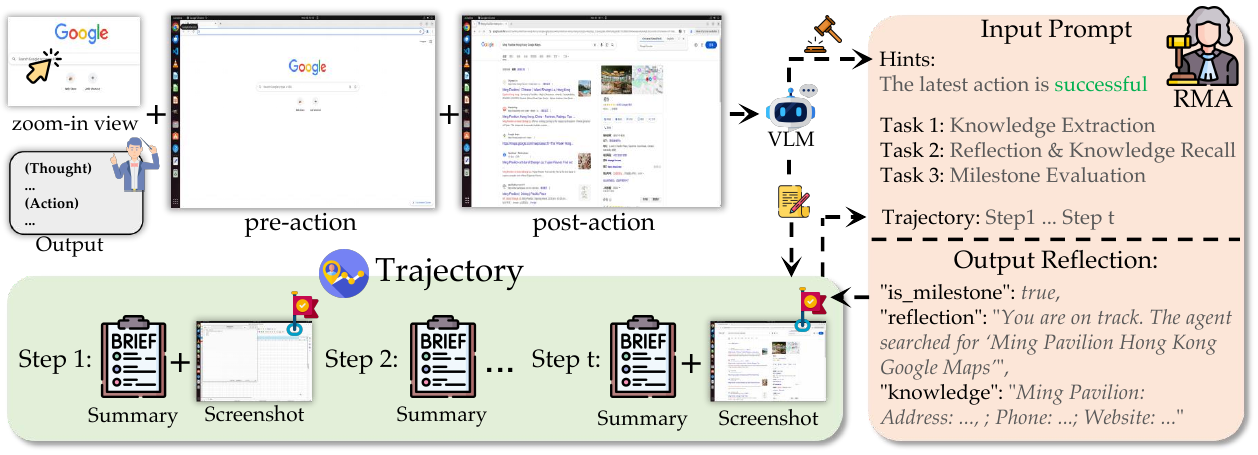} 
    \captionsetup{skip=0pt, position=bottom}
    \caption{\textbf{Pipeline of RMA.} At each step, RMA summarizes the previous action using pre- and post-action screenshots and the Orchestrator’s output, while evaluating the current GUI operation’s correctness. It then generates a reflection from all summaries and milestone screenshots, and determines whether the latest step is a milestone.}
    \label{fig:rma_pipeline}
    \vspace{-0.5cm}
\end{figure*}

Current CUA frameworks suffer from intent drift and insufficient error awareness during long-horizon tasks, due to the lack of a concise yet effective memory mechanism. To address this, we introduce a Reflection-Memory Agent (RMA) which manages a milestone-driven long-term memory to alleviate the Orchestrator's contextual overhead. A crucial insight driving our design is that, despite their information density, screenshots exhibit high temporal redundancy. Consequently, the RMA compresses interaction history while selectively retains only those screenshots identified as milestones. Based on the proposed memory, the RMA generates trajectory-level reflections via a meticulously designed message protocol, thereby providing effective error correction for the Orchestrator. The pipeline of the RMA is shown in Fig.~\ref{fig:rma_pipeline}.

\noindent \textbf{Step-Level Summary.} 
We first utilize an auxiliary VLM to fulfill two tasks: summarizing the latest action and verifying its correctness at the GUI execution level. It can be formally defined as:
\begin{equation}
    \mathcal{S}_{i}, s_i = \mathcal{F}_S(\mathcal{O}_{i-1}, o_{i-1}, o_{i}, \tilde{o}_{i-1}),
    \label{eq:step_summary}
\end{equation}
where $\mathcal{F}_S$ represents the auxiliary VLM, $S_i$ is the execution summary and $s_i$ indicates the success status of the GUI action. $\mathcal{O}_{i-1}$ is the Orchestrator’s output from the previous turn, and $\tilde{o}$ is a localized zoom-in of the action area.

\noindent \textbf{Trajectory-Level Reflection.}
Building upon step-level summaries, we construct a milestone-driven long-term memory module, denoted as $\mathcal{H}_{\text{long}} = \{(\mathcal{S}_j, o_j, m_j)\}_{j=1}^{i-1}$. This module aggregates historical summaries $\mathcal{S}_j$, observations $o_j$, and binary milestone markers $m_j$. Specifically, the marker $m_j$ is generated by the RMA, serving as a gatekeeper such that the RMA processes the observation $o_j$ exclusively when $m_j$ is active (\ie, true).

At each step, the RMA utilizes $\mathcal{H}_{\text{long}}$ to perform three core functions: (1) milestone identification; (2) trajectory-level reflection generation; and (3) relevant information extraction from visual inputs (\eg, retrieving a restaurant detail). 
The formal RMA operation is defined as:
\begin{equation}
    \mathcal{R}_{i}, m_i, k_i = \mathcal{F}_{R}(\mathcal{O}_{i-1}, o_{i},  \mathcal{H}_{\text{long}}),
    \label{eq:rma}
\end{equation}
where $\mathcal{F}_{R}$ represents the RMA and $k_i$ is potential knowledge (or empty if no useful information).

The trajectory-level reflection follows a structured message protocol that categorizes execution states into four classes: \textit{On-track}, \textit{Completed}, \textit{Infeasible}, or \textit{Off-track}. Specifically, we classify \textit{Off-track} scenarios into four distinct error types: (1) GUI Error, where step-level actions fail to achieve expected results (\eg, clicking wrong elements), derived from $s_i$ (Eq.~\ref{eq:step_summary}) as a key heuristic; (2) Lack of Tutorial, triggered when random actions or repetitive loops suggest a need for external guidance; specially, we design a rule-based loop detection algorithm to assist the RMA, and details are provided in the Appendix~\ref{appendix:implementation_detail}; (3) Code Error, identified when a mismatch between the Coder Agent's execution and the actual GUI state; and (4) Other Error, such as intent drift. 

Empowered by this streamlined yet effective memory mechanism, the RMA facilitates high-fidelity reflections for the Orchestrator, which guarantees robust performance across long-horizon tasks. For more details and qualitative case studies, please refer to Appendix~\ref{appendix:implementation_detail} and~\ref{sec:case_study_correct}, respectively.

\subsection{Versatile Tool Agents}

The Orchestrator resolves diverse tasks by synthesizing three specialized tool agents: a novel Searcher to handle out-of-distribution (OOD) knowledge, Grounders for precise UI localization, and a Coder for direct system interface interaction.

\noindent \textbf{Multimodal Searcher.}
Current CUA frameworks still struggle to generalize to OOD tasks. So we introduce a \textit{Visual-Centric Search as a Tool} paradigm, where the Searcher adopts a VLM-driven \textit{SeeAct}~\citep{zheng2024gpt} strategy to interact with rendered pages and synthesize tutorials. Compared to conventional RAG, it preserves critical visual cues beyond text parsing and makes retrieval truly on-demand, invoked only when execution reveals knowledge gaps. The Searcher operates as follows:

(1) First, upon invocation, the Orchestrator formulates a ``\textit{How-to}'' query $q$ and pairs it with the current main-environment observation $o_t$, ensuring retrieval is aligned with the CUA's immediate execution state. To avoid disrupting the main environment, we construct an isolated \textit{Search Environment}, a dedicated browser-based sandbox initialized on the search results page for $q$.

(2) Then, the Searcher operates within a strictly bounded inner loop in the sandbox. We restrict its action space to  $\mathcal{A}_{\text{search}} = \{\texttt{click}, \texttt{type}, \texttt{scroll}\}$, augmented with terminal actions $\{\texttt{done}, \texttt{fail}\}$. This compact design is sufficient for webpage navigation and reading while substantially reducing search complexity. We further encourage the Searcher to visit multiple pages to cross-check and triangulate information.

(3) Finally, the loop terminates under strict criteria. The Searcher returns \texttt{fail} if no relevant information is found within the step budget. Conversely, it triggers \texttt{done} only when the tutorial is deemed highly relevant to $q$. Upon completing the exploration with \texttt{done}, the agent distills the visited content into a structured step-by-step tutorial $\mathcal{T}$, which is permanently appended to the Orchestrator’s context for real-time reference during subsequent steps~(see Appendix~\ref{sec:case_study_correct} for a comprehensive workflow case study).

\noindent \textbf{Grounders.}
For the common UI elements localization, we utilize two complementary agents: (1) a \emph{General Grounder} that integrates low-level visual cues~(\eg, position, appearance) and high-level semantic context~(\eg, functionality, relevance); and (2) an \emph{OCR-based Grounder} tailored for text-intensive apps~(\eg, PowerPoint, Word), which performs word-level OCR to construct a structured \texttt{\{text,id,bbox\}} table, followed by VLM-guided ID selection for precise coordinate lookup. A detailed mapping between actions and grounders is provided in Appendix~\ref{sec:action_space}.

\noindent \textbf{Coder.}
Existing grounding models still struggle with fine-grained localization and exhibit low efficiency in bulk manipulation scenarios. So inspired by \citep{gonzalez2025unreasonable, song2025coact}, we also integrate a Coder specializing in file editing and configuration. At each invocation, the Coder receives a subtask from the Orchestrator, and then starts a strict internal workflow involving file localization, inspection, in-place modification, and verification. After execution, the Coder returns a concise synopsis, and the output is validated by the Orchestrator via GUI state checks. If the Coder fails or validation detects an error, it falls back to completing the subtask via GUI actions.
\section{Experiment}
\begin{table*}[htbp]
  \centering
  \small
  
  \begin{tabularx}{\textwidth}{ l c *{6}{>{\centering\arraybackslash}X} }
    \toprule
    \multirow{2}{*}{\textbf{Method}} & \multirow{2}{*}{\textbf{Step}} 
    & \multicolumn{6}{c}{\textbf{Success Rate}(\%)}  \\
    \cmidrule(lr){3-8}
    & & \textbf{OS} & \textbf{Office} & \textbf{Daily} & \textbf{Prof.} & \textbf{Work.} & \textbf{Avg.} \\
    
    \midrule
    \noalign{\vskip -0.5ex}
    \rowcolor{lightgray} 
    \multicolumn{8}{c}{\textcolor{darkgray}{General Models \& Specialist Model}} \\[-0.5ex]
    \midrule
    Qwen3-VL-32B-Instruct~\citeyearpar{bai2025qwen3} & 50 & -- & -- & -- & -- & -- & 32.40 \\
    Qwen3-VL-32B-Thinking~\citeyearpar{bai2025qwen3} & 50 & -- & -- & -- & -- & -- & 41.00 \\
    
    OpenCUA-72B~\citeyearpar{wang2025opencuaopenfoundationscomputeruse} & 100 & 61.13 & 44.73 & 49.95 & 72.58 & 22.16 & 44.91 \\
    UI-TARS~\citeyearpar{qin2025uitars} & 100 & 41.67 & 50.42 & 55.69 & 51.02 & 14.66 & 41.85 \\
    UI-TARS-2~\citeyearpar{wang2025uiTARS2} & 100 & 41.67 & 61.11 & 62.12 & 61.22 & 34.13 & 53.10 \\
    DeepMiner-Mano-7B~\citeyearpar{fu2025mano} & 100 & 50.00 & 39.28 & 44.87 & 73.47 & 17.20 & 40.15 \\
    DeepMiner-Mano-72B~\citeyearpar{fu2025mano} & 100 & 66.67 & 63.22 & 52.51 & \textbf{83.67} & 24.41 & 53.91 \\
    
    Claude-Sonnet-4.5~\citeyearpar{anthropic2025claude4.5} & 100  & 70.83 & \textbf{72.59} & 61.35 & 63.27 & 49.54 & 62.84 \\
    
    \midrule
    \noalign{\vskip -0.5ex}
    \rowcolor{lightgray}
    \multicolumn{8}{c}{\textcolor{darkgray}{Agentic Framework}} \\[-0.5ex]
    \midrule
    UiPath w/ GPT-5~\citeyearpar{cristescu2025ui} & 50 & 73.91 & 49.52 & 62.12 & 71.43 & 37.30 & 53.69 \\
    CoACT-1 w/ GPT-5~\citeyearpar{song2025coact} & 50 & 70.83 & 60.65 & 54.09 & 69.39 & 42.37 & 56.39 \\
    CoAct-1 w/ GPT-5~\citeyearpar{song2025coact} & 100 & 75.00 & 62.93 & 57.94 & 71.43 & 47.87 & 59.93 \\
    CoAct-1 w/ GPT-5~\citeyearpar{song2025coact} & 150 & 75.00 & 62.93 & 61.78 & 71.43 & 47.87 & 60.76 \\
    
    GTA1 w/ GPT-5~\citeyearpar{yang2025gta1} & 100 & \textbf{79.17} & 63.91 & \underline{62.56} & 79.59 & 50.91 & 63.41 \\
    
    Agent S3 w/ Qwen3-VL-32B-Instruct~\citeyearpar{gonzalez2025unreasonable}$^{\clubsuit}$ & 50 & 50.00 & 36.67 & 50.62 & 61.22 & 21.96 & 40.11 \\
    Agent S3 w/ GPT-5-Mini~\citeyearpar{gonzalez2025unreasonable}$^{\clubsuit}$ & 50 & 62.50 & 54.62 & 46.67 & 44.90 & 37.04 & 47.58  \\
    Agent S3 w/ GPT-5~\citeyearpar{gonzalez2025unreasonable} & 100 & \underline{77.50} & \underline{66.46} & 61.23 & 69.80 & 51.37 & 62.63   \\
    
    \midrule
    
    \name \ w/ Qwen3-VL-32B-Instruct & 50 & 58.33 & 40.94 & 53.54 & \underline{75.10} & 31.24 & 46.86 \\
    \name \ w/ Qwen3-VL-32B-Thinking & 50 & 70.83 & 40.97 & 66.04 & 73.08 & 31.22 & 50.23 \\
    \name \ w/ GPT5-Mini & 50 & 73.68 & 58.17 & 61.39 & 75.00 & 47.37 &  58.05 \\
    \name \ w/ GPT-5 & 50 & 75.00 & 64.85 & 61.19 & 69.23 & \underline{54.86} & \underline{63.61} \\
    \name \ w/ GPT-5 & 100 & \textbf{79.17} & 65.73 & \textbf{67.76} & 69.23 & \textbf{57.98} & \textbf{65.84} \\
    \bottomrule
  \end{tabularx}
  
  \captionsetup{skip=3pt, position=bottom}
  \caption{Main results of \name\ on OSWorld. $^{\clubsuit}$ represents the result reproduced by us, and others are sourced from the official leaderboard. Bold highlights the best performance, and underline denotes the runner-up.}
  \label{tab:main_results}
  
  \vspace{-0.4cm}
\end{table*}

\subsection{Experiment Setup}
\noindent \textbf{Evaluation Benchmarks.}
Our evaluation centers on desktop environments, employing \textbf{OSWorld-Verified}\footnote{In this paper, ``OSWorld'' refers to ``OSWorld-Verified''.}~\citep{xie2024osworld,osworld_verified} as our primary benchmark, which comprises 369 real-world tasks across five domains in a Ubuntu environment. Following common practice~\citep{xie2024osworld}, we exclude the 8 Google Drive tasks, yielding a final set of 361 tasks. To further probe cross-platform generalization, we incorporate WindowsAgentArena~\citep{bonatti2024windows} and MacOSArena~\citep{wang2025mmbenchgui} (see Appendix~\ref{sec:extra_results_win_mac}). As extensions of the OSWorld paradigm, these benchmarks assess framework robustness in both cross-platform software consistency and system-specific configurations.

\noindent \textbf{Baselines.}
We compare \name~against leading general models, specialist native agents and agentic frameworks, highlighting Agent S3~\citep{gonzalez2025unreasonable} and CoAct-1~\citep{song2025coact} as primary baselines, as they share a similar action space, characterized by the utilization of a coder for task execution.

\noindent \textbf{Implementation Details.}
We select distinct based VLMs, encompassing both proprietary and open-source models across varying capability levels: GPT-5~\citep{openai2025gpt5systemcard}, GPT-5-Mini~\citep{openai2025gpt5systemcard} and Qwen3-VL series~\citep{bai2025qwen3}. In our main experiments, we employ a single VLM to drive all core roles within \name, including Orchestrator, RMA, Searcher, and Coder.
For grounding, we leverage UI-TARS-1.5-7B~\citep{qin2025uitars} as the General Grounder and EasyOCR\footnote{\url{https://github.com/JaidedAI/EasyOCR}} as the OCR Grounder. Regarding hyperparameters, we set the VLM temperature to 0.1 to maximize generation stability, with a maximum context window of 8 \texttt{1920x1080} images. Additionally, the EasyOCR width threshold is configured to 0.1 to ensure precise word-level recognition.

\subsection{Main Results}

\noindent \textbf{OSWorld.}
As shown in Tab.~\ref{tab:main_results}, \name\ with GPT-5 establishes a new SOTA with 63.61\% (50-step) and 65.84\% (100-step), surpassing the primary baseline (100-step Agent S3 w/ GPT-5) by $\sim$3\%, demonstrating the effectiveness of our framework. Notably, in the \textit{Workflow} domain, which involves interactions across multiple applications, \name\ achieves a more substantial gain, outperforming the runner-up Agent S3 by 7\%. 
This strong performance in long-horizon tasks is largely attributable to the RMA. With a streamlined yet robust design, the RMA acts as a critical safeguard against error accumulation, promoting long-term temporal stability and overall system resilience.

Furthermore, we reproduced Agent S3 using Qwen3-VL-32B-Instruct and GPT-5-Mini for a direct comparison. The results indicate that \name\ yields a gain of 8\% with Qwen3-VL, with the improvement expanding to 10\% when using GPT-5-Mini. Notably, these gains are more pronounced in models with relatively lower reasoning capacities. We attribute this to our \textit{Visual-Centric Search as a Tool} design, which effectively compensates for the knowledge deficits inherent in smaller models. For instance, under a 50-step limit, the GPT-5-Mini variant invoked the Searcher 34 times more frequently than its GPT-5 counterpart. While stronger models leverage vast parametric knowledge to solve tasks directly, weaker models rely on our framework's search capabilities to bridge the information gap.

These results also highlight the exceptional cost-effectiveness of our approach. Specifically, \name\ leveraging GPT-5-Mini exhibits a marginal performance delta of only 3\% compared to Agent S3 (driven by the significantly more powerful GPT-5), thereby attaining competitive efficacy with substantial cost reduction. Besides, when applied to open-source models, \name\ yields remarkable improvements: it achieves relative improvements of 45\% and 23\% for Qwen3-VL-32B-Instruct and Qwen3-VL-32B-Thinking, respectively, over their vanilla counterparts. This underscores a critical insight: \name\ not only establishes new SOTA standards but also democratizes advanced agentic capabilities, enabling smaller models to deliver competitive performance through a cohesive framework.

\begin{table}[t]
  \centering
  \small
  
  \begin{tabularx}{\linewidth}{ l c c }
    \toprule
    \textbf{Method} & \textbf{Step} & \textbf{Avg.}(\%) \\
    
    \midrule
    Qwen3-VL-32B-Instruct~\citeyearpar{bai2025qwen3} & 50 & 31.7 \\
    UI-TARS-1.5-7B~\citeyearpar{qin2025uitars} & 50 & 42.1 \\
    UI-TARS-2~\citeyearpar{wang2025uiTARS2} & 50 & 50.6 \\
    Agent S3 w/ GPT-5~\citeyearpar{gonzalez2025unreasonable} & 50 & 54.1 \\
    Agent S3 w/ GPT-5~\citeyearpar{gonzalez2025unreasonable} & 100 & 56.6 \\
    
    \midrule
    
    \name \ w/ Qwen3-VL-32B-I. & 50 & 45.3 \\
    \name \ w/ GPT-5-Mini & 50 & \underline{62.2} \\
    \name \ w/ GPT-5 & 50 & \textbf{63.5} \\ 
    
    \bottomrule
  \end{tabularx}
  \captionsetup{skip=3pt, position=bottom}
  \caption{Main results of \name\ on WindowsAgentArena.}
  \label{tab:main_results_waa}
  \vspace{-0.5cm}
\end{table}
\noindent \textbf{WindowsAgentArena.}
As shown in Tab.~\ref{tab:main_results_waa}, \name\ establishes a new SOTA on WindowsAgentArena, achieving 63.5\% with GPT-5 under a 50-step limit. This performance surpasses the 50-step and 100-step Agent S3 baselines by 9.4\% and 6.9\%, respectively. Remarkably, even the GPT-5-Mini variant demonstrates superior efficiency, exceeding the 100-step Agent S3 (GPT-5) by 5.6\%. Furthermore, the framework exhibits strong robust adaptability across model scales. Specifically, when configured with Qwen3-VL-32B-Instruct, \name\ attains a success rate of 45.3\%; while this trails the specialist UI-TARS-2, it represents a substantial 13.6\% improvement over the vanilla baseline. These results indicate that our tailored designs not only accommodate models of varying scales but also enable effective generalization to distinct OS-level characteristics. Detailed results and analysis are provided in Appendix~\ref{sec:extra_results_win_mac}.

\begin{table}[t]
  \centering
  \resizebox{\columnwidth}{!}{
  
  \begin{tabular}{l | ccc | cc}
    \toprule
    \multirow{2}{*}{\textbf{Method}} 
    
    & \multicolumn{3}{c|}{\textbf{Success Rate}(\%)}

    & \multirow{2}{*}{\textbf{Token(k)}}
    & \multirow{2}{*}{\textbf{Step}} \\
    \cmidrule(lr){2-4}
    
    & \textbf{Daily} & \textbf{Workflow} & \textbf{Avg.} &  & \\
    
    \midrule
    \noalign{\vskip -0.5ex}
    \rowcolor{lightgray}
    
    \multicolumn{6}{c}{\textcolor{darkgray}{Combination Ablation}} \\[-0.5ex] 
    \midrule

    w/o Search, Refl. & 49.60 & 37.33 & 51.90 & 471 & 18.4 \\
    
    \midrule
    \noalign{\vskip -0.5ex}
    \rowcolor{lightgray} 
    \multicolumn{6}{c}{\textcolor{darkgray}{Search Ablation}} \\[-0.5ex] 
    \midrule
    w/o Search & 50.65 & \textbf{48.06} & 53.78 & 801 & 14.7 \\
    +) Unimodal & 56.10 & 39.30 & 54.81 & 900 & 15.2  \\
    +) Multimodal (Ours) & \textbf{61.86} & 47.37 & \textbf{58.05} & 900 & 15.2 \\
    
    \midrule
    \noalign{\vskip -0.5ex}
    \rowcolor{lightgray}
    \multicolumn{6}{c}{\textcolor{darkgray}{Reflection \& Memory Ablation}} \\[-0.5ex] 
    \midrule
    w/o Reflection & 60.20 & 39.23 & 54.38 & 535 & 18.5 \\
    +) Refl. w/ STM. & 56.10 & 39.49 & 54.01 & 843 & 14.9  \\
    +) Refl. w/ LTM.(Ours) & \textbf{61.86} & \textbf{47.37} & \textbf{58.05} & 900 & 15.2 \\
    \bottomrule
  \end{tabular}
  }
  \captionsetup{skip=3pt, position=bottom}
  \caption{Ablation study on OSWorld. Experiments utilize GPT-5-Mini and UI-TARS-1.5-7B with a 50-step limit. STM: Short-Term Memory (``Last-K turns''); LTM: Long-Term Memory (our method).}
  \label{tab:ablation_results}
  \vspace{-0.5cm}
\end{table}

\subsection{Ablation Study}

In this section, we conduct a comprehensive ablation analysis focusing on the core contributions of our framework: the Searcher and RMA, with quantitative results presented in Tab.~\ref{tab:ablation_results}. 

For Unimodal Search, we adopt an inner-loop search paradigm consistent with our Multimodal Search, utilizing \texttt{search} and \texttt{parse} tools via SearXNG\footnote{\url{https://github.com/searxng/searxng}} and Crawl4AI~\citep{crawl4ai2024}. Our approach consistently outperforms all ablation baselines overall. 
Notably, the \textit{Daily} domain, which inherently demands extensive external knowledge, benefits significantly from the search module.
Specifically, Multimodal Search achieves substantial relative gains of 22.1\% and 10.3\% over the w/o Search and Unimodal Search baselines, respectively. Further, manual trajectory inspection confirmed that Multimodal Search performance aligns with expectations.

The \textit{Workflow} domain typically involves cross-application, long-horizon tasks that demand robust long-term memory storage and comprehension, rendering the RMA module pivotal. Compared to the \textit{Refl. w/ STM} paradigm and the \textit{w/o Reflection} setting, our RMA delivers substantial relative improvements of approximately 20.0\% and 20.7\%, respectively. 
This result validates the efficacy of abstract trajectory memory with granular error reflection in handling complex interactions. Notably, the \textit{w/o Reflection} baseline marginally outperforms \textit{Refl. w/ STM} while reducing token consumption by $\sim$36.5\%. This suggests that naive Last-K reflection may be ineffective or even detrimental, implying that omitting reflection entirely is preferable to deploying a suboptimal implementation.

Finally, regarding step efficiency, we observe that ablating RMA results in an increase of 3.3 steps per task. This underscores the reflection module's capacity for timely error identification and rectification, thereby preventing futile exploration and streamlining task completion.

\subsection{Discussion}
\label{sec:discussion}
\begin{table}[t]
  \centering
  
  \small
  \resizebox{\columnwidth}{!}{
  
  \begin{tabular}{l | cc | c}
    \toprule
    \multirow{2}{*}{\textbf{Model}}
    
    & \multicolumn{2}{c|}{\textbf{Success Rate}(\%)} 
    & \multirow{2}{*}{\textbf{Cost(\$)}} \\
    \cmidrule(lr){2-3}
    
    & \textbf{Workflow} & \textbf{Avg.} & \\
    
    \midrule
    \noalign{\vskip -0.5ex}
    \rowcolor{lightgray}
    
    \multicolumn{4}{c}{\textcolor{darkgray}{Different Based VLMs w/ UI-TARS-1.5-7B}} \\ [-0.5ex] 
    \midrule

    Claude-Sonnet-4.5~\citeyearpar{anthropic2025claude4.5} & \textbf{57.21} & - & $\approx500$ \\
    GPT-5 & 54.86 & 63.61 & $\approx150$ \\
    GPT-5-Mini & 47.37 & 58.05 & $\approx30$ \\
    Qwen3-VL-32B-Instruct & 31.24 & 46.86 & $0$ \\    
    
    \midrule
    \noalign{\vskip -0.5ex}
    \rowcolor{lightgray} 
    
    \multicolumn{4}{c}{\textcolor{darkgray}{Different Grounders w/ Qwen3-VL-32B-Instruct}} \\ [-0.5ex] 
    \midrule
    
    UI-TARS-1.5-7B & 31.24 & 46.86 & $0$ \\
    GTA1-32B~\citeyearpar{yang2025gta1} & 28.54 & 46.32 & $0$ \\
    ScaleCUA-32B~\citeyearpar{liu2025scalecua} & 31.76 & 45.76 & $0$ \\
    Holo2-30B-A3B~\citeyearpar{hai2025holo2modelfamily} & 24.45 & 43.92 & $0$ \\
    \bottomrule
  \end{tabular}
  }
  
  \captionsetup{skip=3pt, position=bottom}
  \caption{Impact of based VLMs and grounders configurations on OSWorld (50-step limit). `Cost' represents the total expenditure for the \textit{Workflow} domain, where \$0 denotes local deployment of open-source models.}
  \label{tab:backbone_grounder_results}

  \vspace{-0.5cm}
\end{table}

\noindent \textbf{Impact of Based VLMs and Grounders.}
This section analyzes the sensitivity of our framework to different based VLMs and grounding models. 
Note that Claude-Sonnet-4.5 was tested only on the \textit{Workflow} domain due to high inference costs.
Tab.~\ref{tab:backbone_grounder_results} reveals a strong correlation between model scale and performance. Claude-Sonnet-4.5 achieves the highest \textit{Workflow} score~(57.21\%) but comes with a steep price tag~($\sim$\$500). In contrast, cost-efficient GPT-5-Mini offers a compelling balance: it trails GPT-5 by only 5\% on average while reducing costs by $\sim$80\%. 
On the other hand, the performance remains stable across different grounding models, underscoring the robustness of \name. This efficacy is rooted in our collaborative architecture, which delegates fine-grained tasks to the Coder, thereby mitigating dependency on the Grounder’s specific proficiency.

\noindent \textbf{Key Insights.}
Finally, we present three key insights derived from our experiments:

\textbf{The Granularity Gap in Visual Perception.} While generally robust, the RMA falters against subtle visual nuances. Current VLMs often fail to resolve fine-grained cues like highlighting or overlapping windows. This \textit{perceptual blindness} leads the RMA to issue false positive errors, paradoxically allowing the \textit{w/o RMA} baseline to outperform the full framework in visually complex domains~(see Appendix~\ref{sec:error_case} for illustrative cases). Future advancements must pivot towards targeted prompt engineering or enhanced image post-processing to bridge this granularity gap.

\textbf{Bottlenecks in the Planner-Worker Paradigm.} Precise textual articulation of visual affordances (\eg, ``boundaries'') is challenging, causing information bottlenecks between the Orchestrator and Grounder. We argue that end-to-end native CUAs are essential to resolve this by eliminating textual abstraction. Future work should pivot towards hybrid paradigms, integrating native CUA capabilities to break through the ceiling of current paradigm.

\textbf{Stochasticity: Volatility \vs Latent Potential.} Despite strict control over prompts and temperature, we observe significant task-level volatility. While detrimental to deployment stability, this variance masks high latent competence: aggregating results via Pass@5 boosts performance to 79.40\%, significantly surpassing the human baseline (72.4\%). This suggests that the model \textit{can} solve the tasks but lacks consistency. Harnessing this ``volatility'' 
remains a critical frontier for future framework design.

\section{Conclusion}

In this work, we present \name, a holistic Computer-Using Agent (CUA) framework containing an Orchestrator that synergistically coordinates specialized modules to address the two challenges of long-horizon robustness and domain generalization. To mitigate the visual loss inherent in extended workflows, our Reflection-Memory Agent employs a milestone-driven strategy for granular memory curation, enabling retrospective auditing to rectify errors such as intent drift and loops. Besides, we employ the Versatile Tool Agents featuring a \textit{Multimodal Searcher} that transcends text-based retrieval limitations by adopting an active \textit{SeeAct} paradigm, synthesizing high-fidelity, visually aligned tutorials for unseen environments. Extensive experiments confirm that \name\ not only achieves state-of-the-art performance across diverse operating systems but also proves that complex problems can be effectively solved using open-source VLMs. We hope our insights will provide a resilient blueprint for future real-world CUAs. 
\section*{Limitations}

Despite the robust performance of \name\ across major desktop environments, several limitations remain. 

\noindent \textbf{Environmental Generalization.} Our current evaluation is strictly confined to desktop ecosystems. The framework’s adaptability to mobile platforms, such as Android and iOS, remains unverified due to the necessity of distinct action space adaptations for mobile interfaces. Consequently, achieving full cross-platform universality remains a subject for future exploration.

\noindent \textbf{Structural Complexity and Efficiency.} The multi-agents system introduces inherent overhead. Although we employ strategies to mitigate error accumulation, the extensive inter-agent interactions result in high token consumption and significant latency. Specifically, our execution speed is tens of times slower than human performance, which currently precludes real-time deployment. Future iterations may address this via dynamic "fast and slow" reasoning mechanisms or simplified architectures.

\ifsubmission
\else
   \noindent \textbf{Implementation Specifics.} Our memory mechanism currently relies on established summarization paradigms, and the searcher module may eventually be superseded by more robust commercial engines (\eg, Google AI Search) when their integration overhead becomes acceptable. While these specific sub-components represent current implementation choices, future work could focus on upgrading them within our collaborative framework, which is designed to adapt to and orchestrate more advanced tools as they emerge.
\fi

\section*{Ethical Considerations}
The development of autonomous Computer-Using Agents~(CUAs) introduces significant ethical and safety responsibilities. In this work, we prioritized operational safety by conducting all evaluations within strictly isolated, sandboxed virtual environments (\eg, Docker containers and virtual machines). This isolation ensures that the agent's exploratory actions can't inadvertently damage host systems or access unauthorized external networks during the research phase.

However, transitioning from controlled benchmarks to real-world deployment necessitates a rigorous re-evaluation of system security and user privacy. Since visual agents like \name\ inherently process continuous streams of screenshots, they possess the capability to ``see'' sensitive personal identifiable information displayed on the screen. Consequently, deploying such agents requires the implementation of strict, granular permission controls and robust data sanitization protocols. Users must retain absolute authority to define the agent's operational boundaries, ensuring that sensitive applications (\eg, banking, private messaging) remain inaccessible unless explicitly authorized.

Furthermore, the robust automation capabilities demonstrated by our framework carry inherent dual-use risks. While designed to enhance human productivity, these systems could potentially be exploited for auto-execution of malicious workflows if not properly safeguarded. We posit that the advancement of CUAs' capabilities must proceed in lockstep with the development of ``Safety by Design'' principles. Future research must focus not only on increasing success rates but also on embedding deep alignment mechanisms, ensuring that agents remain reliable, transparent, and strictly adherent to human ethical standards in open-ended digital environments.

\bibliography{custom}

\newpage
\appendix

\section{Details of \name}

\label{sec:detail}
In this section, we provide a comprehensive elaboration on the implementation details of our \name \ framework to facilitate a deeper understanding. So, how exactly is this \textbf{Symphony} composed?

\subsection{Task Definition}

The interaction process of Computer-Using Agents~(CUAs) can be modeled as a Partially Observable Markov Decision Process (POMDP), defined by the tuple $(S, A, O, T, \mathcal{O})$. Here, $S$ represents the set of environmental states; $A$ denotes the finite set of executable actions available to the agent; and $O$ represents the set of observations the agent can receive. The state transition function $T(S'|S, A) \rightarrow [0,1]$ defines the probability of transitioning to a new state given the current state and the action. The observation function $\mathcal{O}(O|S, A) \rightarrow [0,1]$ defines the probability of receiving a specific observation given a state and action. For brevity, the discount factor and the reward function are omitted in this formulation.
Addressing the partial observability, the state of a GUI environment consists of a complex composition of numerous GUI elements, resulting in an infinite state space. However, for a purely vision-based approach, the observation of the GUI environment  is a screenshot. For an RGB screenshot with dimensions $H \times W$, the possible observations are limited to $H \times W \times 3$ variations. Consequently, there is an inherent information compression from the state space to the observation space. The following discussion focuses on the observation $o$.

To formalize the task, we introduce the task goal (user instruction) $\mathcal{I}$, where the agent receives observations from the environment and executes corresponding actions to complete the task. To further enhance the agent's reasoning capabilities and support more complex decision-making processes~\citep{yao2023react}, a "thought" component $t_i$ is incorporated prior to each action $a_i$. This thought process serves as a critical intermediate step that analyzes the current situation and historical actions, ensuring the rationality and intentionality of each decision. Consequently, the entire task process can be formalized as a trajectory:
\begin{equation}
\mathcal{I}, (o_1, t_1, a_1), (o_2, t_2, a_2), \dots, (o_n, t_n, a_n)
\label{eq:trajectory_thought}
\end{equation}

In a POMDP, the agent can only infer the current true state through observations; thus, a single observation is insufficient to fully depict the environmental state. To address this, an implicit Belief State can be constructed by aggregating historical information to more accurately approximate the current true state. The update process of this belief state is essentially an extension of the Markov property. Although a single observation may fail to satisfy Markovian conditions, the aggregation of historical information allows the belief state to be viewed as adhering to Markov properties.

In an ideal CUA architecture, to achieve optimal decision-making, the model should theoretically access and process the entire interaction history from the inception of the task. This complete historical record contains all contextual cues regarding environmental evolution and serves as the foundation for constructing a precise belief state. Consequently, conditioned on the task instruction and the full history preceding the current step, the model iteratively predicts the thought $t_i$ and action $a_i$ until the action $\{\texttt{done}, \texttt{fail}\}$ appears. This probabilistic model is formalized as:
\begin{equation}
P(t_i, a_i | \mathcal{I}, (o_{j}, t_{j}, a_{j})_{j=1}^{i-1}, o_i)
\label{eq:prediction_prob_full}
\end{equation}

While theoretically optimal, the complete history model in Eq.~\ref{eq:prediction_prob_full} faces practical limitations in long-horizon GUI tasks. Despite extended LLM contexts ($\geq128K$), directly incorporating full multimodal histories incurs significant drawbacks: (1) \textbf{Efficiency and Reliability Issues}: Processing dozens of screenshots and textual histories increases computational overhead and may induce hallucinations due to information overload; (2) \textbf{Historical Redundancy}: Many intermediate observations (\eg, failed attempts or trivial clicks) provide diminishing informational value for future decisions.

Thus, effective \textbf{context compression} becomes crucial—filtering redundant content while preserving decision-critical information. We formalize this as finding an optimal compression function $\mathcal{C}$ that transforms raw history $\mathcal{H}_{1:i-1}$ into condensed representation $\tilde{\mathcal{H}}_{1:i-1}$, maximizing the likelihood of correct actions:
\begin{equation}
\max_{\mathcal{C}} \mathbb{E} \left[ P\left( t_i^*, a_i^* \mid \mathcal{I}, \mathcal{C}(\mathcal{H}_{1:i-1}), o_i \right) \right]
\end{equation}

This yields the practical decision model:
\begin{equation}
P(t_i, a_i \mid \mathcal{I}, \tilde{\mathcal{H}}_{1:i-1}, o_i)
\label{eq:summary}
\end{equation}

\begin{table*}[t]
  \centering
  \footnotesize 
  \setlength{\tabcolsep}{4pt}

  \begin{tabularx}{\textwidth}{
    >{\RaggedRight}p{2.0cm}          
    >{\RaggedRight\hsize=1.1\hsize}X 
    >{\RaggedRight}p{2.2cm}          
    >{\RaggedRight\hsize=0.9\hsize}X 
    l                                
    l                                
  }
    \toprule
    \textbf{Error Type} & \textbf{Description} & \textbf{Aux. Detection} & \textbf{Method Desc.} & \textbf{Type} & \textbf{Rel. Tool} \\
    \midrule
    
    GUI Error & 
    Failures at the execution level where the intended action (e.g., click, type) is unsuccessful. & 
    Step Summary & 
    A low-level analysis correlating actions with pre- and post-transition screenshots. &
    VLM &
    Grounder \\
    \addlinespace
    \midrule
    
    Lack of Tutorial & 
    The agent performs technically correct operations but lacks procedural logic to advance the workflow. & 
    Loop Detection & 
    A loop detection algorithm evaluating similarity metrics across sequential actions and visual states. &
    Rule &
    Searcher \\
    \addlinespace
    \midrule
    
    Code Error &
    Post-execution discrepancies identified during verification. Output fails to align with instructions. & 
    -- & 
    -- &
    -- &
    -- \\
    \midrule

    Other Error &
    Deviations not covered by other categories, such as drifting from the primary goal, factual errors, or hallucinations. & 
    -- & 
    -- &
    -- &
    -- \\
    
    \bottomrule
  \end{tabularx}

  \captionsetup{skip=3pt, position=bottom}
  \caption{Definitions of Error Types, Auxiliary Methods, and Relevant Tools.}
  \label{tab:error_types}
  \vspace{-0.3cm}
\end{table*}

The design of $\mathcal{C}$ is therefore central to enabling efficient long-horizon GUI task performance. In this paper, we dedicate our research to this pivotal challenge, presenting in-depth explorations into the optimization of such compression mechanisms.

\subsection{More Details of \name}
\label{appendix:implementation_detail}

We provide further implementation details of the modules that were omitted in Section~\ref{sec:method} due to space constraints. The following paragraphs elaborate on the Orchestrator, Reflection-Memory Agent, General Grounder, OCR Grounder, Searcher, and Coder.

\paragraph{Orchestrator.}
Serving as the cognitive core of \name, the Orchestrator manages short-term memory to iteratively generate actions, as defined in Eq.~\ref{eq:summary}. To balance context retention with efficiency, it implements the compression function $\mathcal{C}$ by synthesizing a sliding window of recent interactions with high-level semantic insights. Formally, the condensed history is constructed as:
\begin{equation}
\tilde{\mathcal{H}}_{1:i-1} = \left\{ (o_{j}, t_{j}, a_{j}) \right\}_{j=i-K+1}^{i-1} \cup \{ \mathcal{R}_{i} \},
\end{equation}
where the context combines the raw trajectory of the most recent $K$ steps with reflections on the immediate previous execution and the retrieval of relevant procedural knowledge for the current step provided by the Reflection-Memory Agent ($\mathcal{R}_{i}$). This hybrid memory structure allows the Orchestrator to maintain minimal context length without sacrificing the critical semantic cues necessary for precise decision-making.

\paragraph{Reflection-Memory Agent.}

Serving as a key component of our symphony and specifically engineering for long-term memory management, RMA is designed to provide precise feedback for the subsequent step. It distills the current trajectory into an abstract representation comprising multiple milestone screenshots and a comprehensive history of step-wise transitions. Furthermore, it extracts critical memories to construct a procedural knowledge base. In this subsection, we further elaborate on the two corresponding auxiliary detection methods with defined core error types, as shown in Tab.~\ref{tab:error_types}.

\textbf{Step Summary.} Before the Orchestrator's decision making process based on the current observation $o_t$, we execute a retrospective analysis of the last interaction, formalized as Eq.~\ref{eq:step_summary}. To bolster the VLM's perception, we introduce $\tilde{o}_{i-1}$, a ``zoom-in'' augmentation applied solely when $a_{i-1}$ involves coordinates; this consists of a $400$-pixel radius crop centered on the element's coordinate, overlaid with a eye-catching red visual marker. 

Notably, we refrain from augmenting $o_i$ similarly, as the visual consequences of a GUI interaction often manifest in regions distinct from the initial actuation coordinates. Ultimately, this low-level verification focuses on single-step fidelity, providing $s_{i}$ as a critical input for the step behavior history and a decisive signal for RMA's detection of \textit{GUI Error}.

\textbf{Loop Detection.}
While chaotic and disorganized trajectory states are difficult to probe through rule-based methods, repetitive and cyclical trajectories are easily perceptible. Therefore, we designed a similarity-rule-based loop detection algorithm:
\begin{equation}
    \small
    D_{loop}(H, N) = \bigvee_{k=T-2N}^{1} \bigwedge_{j=0}^{N-1} \mathcal{M}(k+j, T-N+j),
    \label{eq:loop_detection}
\end{equation}
where $\mathcal{M}(u, v) \triangleq \mathcal{S}_{img}(o_u, o_v) \land \mathcal{S}_{act}(a_u, a_v)$ denotes the joint similarity check for observation $o$ and action $a$ at time steps $u$ and $v$. $H$ represents the trajectory history, and $N$ is the sliding window size (default set to 3). In practice, we scan $k$ in reverse order (from $T-2N$ to $1$) with early stopping: once $\bigwedge_{j=0}^{N-1}\mathcal{M}(k+j,\,T-N+j)$ holds, we terminate and return the most recent matched segment.

To implement the metric defined in Eq.~\ref{eq:loop_detection}, we prioritize high precision to ensure that feedback provided to the RMA is factually grounded, adopting a strict ``coarse-to-fine" matching protocol. For visual state similarity $\mathcal{S}_{img}$, we employ a cascaded verification strategy: a Perceptual Hashing (pHash)~\citep{zauner2010phash} check first filters distinct states using a tight Hamming distance threshold ($\le 1$), followed by a Structural Similarity Index (SSIM)~\citep{wang2004ssim} calculation with a high threshold ($0.99$) to confirm identity despite minor rendering artifacts. Action similarity $\mathcal{S}_{act}$ is determined based on action semantics: coordinate-dependent actions (\eg, \textit{click}, \textit{scroll}) require Euclidean distances within a relative threshold (5\% of the screen diagonal) alongside matching discrete parameters; discrete actions (\eg, \textit{type}, \textit{open}) demand exact argument matching; and natural language queries (\eg, \textit{search}) utilize Levenshtein distance with a high similarity threshold to tolerate minor phrasing variations. The search process iterates backwards from $k=T-2N$ to identify the most recent historical interval $[k, k+N-1]$ identical to the current window. To ensure real-time performance, we adopt a space-for-time optimization strategy by caching image features, effectively reducing the computational complexity from $O(T \cdot N \cdot C_{img})$ to $O(T \cdot (N+C_{img}))$ by avoiding redundant image processing during the sliding window comparison, $C_{img}$ represents the computational complexity of pHash and SSIM algorithm between two images. The loop detection algorithm prioritizes high precision, a successful loop identification provides a strong signal for RMA's detection of \textit{Lack of Tutorial}.

\paragraph{General Grounder.}

Serving as the visual grounding engine, this VLM-based Grounder localizes UI elements by processing hybrid descriptions that integrate low-level visual cues (position, appearance) with high-level semantic context (functionality, instruction relevance). Based on our empirical evaluation of existing grounding models including UI-TARS-1.5-7B~\citep{qin2025uitars}, ScaleCUA-32B~\citep{liu2025scalecua}, Holo1.5-72B~\footnote{\url{https://www.hcompany.ai/blog/holo2}}, Holo2-30B-A3B~\citep{hai2025holo2modelfamily}, GTA1-32B~\citep{yang2025gta1} and GroundNext-7B~\citep{feizi2025groundcua}. UI-TARS-1.5-7B demonstrates the best performance among open-source models in desktop environments, followed by GTA1-32B and ScaleCUA-32B.

\paragraph{OCR-based Grounder.}

Serving as a complementary complement to the General Grounder, the OCR-based Grounder operates in synergy with the VLM Grounder to enhance element localization, aiming to address the General Grounder's limitations in resolving precise word-boundary coordinates. The workflow entails a word-level OCR scan that generates a structured table of \texttt{\{text, id, bbox\}}. This table is subsequently processed by the general VLM for semantic ID selection, enabling precise coordinate retrieval via index lookup. This approach effectively mitigates performance deficits in text-dense domains such as PowerPoint and Word. However, it represents a pragmatic compromise because current OCR models still lack the granularity for character-level localization, a capability not yet strictly demanded by existing benchmarks. Currently, resolving this limitation necessitates either using code tools for intrinsic file manipulation or rewriting large text blocks as a shortcut to bypass precise localization. 

\paragraph{Searcher.}

Serving as the core module for external knowledge retrieval, the Searcher employs a visual browsing strategy to navigate the open web and synthesize step-by-step tutorials. Crucially, we enforce a strict validity constraint: the agent is instructed to return a tutorial only when high relevance is guaranteed, defaulting to a \texttt{fail} state otherwise. This design prevents the contamination of the Orchestrator's context with lengthy or erroneous information that could disrupt downstream decision-making.

\paragraph{Coder.}

Serving as the core execution module for system-level tasks, the Coder interacts directly with the environmental CLI to execute Shell and Python code, excelling in file revision and configuration. To mitigate complexity, the Orchestrator delegates sub-tasks to the Coder, which follows a strict internal workflow involving file localization, content inspection, in-place modification via complete overwrites, verification and visualization.
Following execution, a summary agent module provides an execution synopsis to the Orchestrator as a textual observation. 
While the Coder's GUI-free design ensures high efficiency during iterative processes, the resulting file modifications can manifest as abrupt GUI mutations imposing a high cognitive load, thereby necessitating a unified verification protocol across all agents (Coder, RMA, Orchestrator) to minimize side effects deviating from user instructions.

Ultimately, the designs of both the Searcher and Coder embody the principle of \textit{Context Folding}. This paradigm entails offloading independent sub-tasks to isolated execution contexts to prevent polluting the Orchestrator's context window. By ``folding'' the detailed execution trajectory and results into a concise summary, we maintain a seamless logical flow for the Orchestrator. This encapsulation strategy offers a generalizable method for managing context in complex agentic systems.

\begin{table*}[tbp]
  \centering

  \begin{tabularx}{\textwidth}{l l X}
    \toprule
    \textbf{Action} & \textbf{Category} & \textbf{Parameter Specification} \\
    \midrule

    click & GUI (General Grounding) & 
    \textbf{Format:} \texttt{[desc,num\_clicks,button,hold\_keys]}\newline
    \textbf{Details:} Target element description; clicks number; button to click; keys to hold. \\
    \addlinespace
    
    type & GUI (General Grounding) &
    \textbf{Format:} \texttt{[desc,text,overwrite,enter,terminal]}\newline
    \textbf{Details:} Target element description; text content; overwrite flag(bool); press enter after typing(bool); terminal flag(bool). \\
    \addlinespace
    
    scroll & GUI (General Grounding) &
    \textbf{Format:} \texttt{[desc,clicks,shift]}\newline
    \textbf{Details:} Target element description; clicks (+up/-down); shift for horizontal scroll(bool). \\
    \addlinespace
    
    drag\_and\_drop & GUI (General Grounding) &
    \textbf{Format:} \texttt{[start\_desc,end\_desc,hold\_keys]}\newline
    \textbf{Details:} Descriptions for start/end locations; keys to hold during drag. \\
    \addlinespace
    
    highlight\_text\_span & GUI (OCR Grounding) &
    \textbf{Format:} \texttt{[start\_phrase,end\_phrase,button]}\newline
    \textbf{Details:} Unique anchor phrase; button to hold. \\
    \addlinespace
    
    locate\_cursor & GUI (OCR Grounding) &
    \textbf{Format:} \texttt{[phrase,pos,text]}\newline
    \textbf{Details:} Unique anchor phrase; position(start/end); optional text to insert immediately. \\
    \addlinespace
    
    hotkey & GUI (No Grounding) &
    \textbf{Format:} \texttt{[keys]}\newline
    \textbf{Details:} List of keys to press in combination (e.g., [`ctrl', `c']). \\    
    \addlinespace
    
    hold\_and\_press & GUI (No Grounding) &
    \textbf{Format:} \texttt{[hold\_keys,press\_keys]}\newline
    \textbf{Details:} Keys to hold down while pressing a sequence of other keys. \\
    \addlinespace
    
    open & GUI (No Grounding) &
    \textbf{Format:} \texttt{[app\_or\_filename]}\newline
    \textbf{Details:} Name of the application or file to open. \\

    \midrule
    
    call\_search\_agent & Proprietary & 
    \textbf{Format:} \texttt{[query]}\newline
    \textbf{Details:} A "How to" question targeting a specific tutorial (e.g., "How to apply filters in Excel?"). \\
    \addlinespace
    
    call\_code\_agent & Proprietary & 
    \textbf{Format:} \texttt{[task]}\newline
    \textbf{Details:} A self-contained goal executable via code (e.g., data analysis, file processing). \\

    \midrule
    
    wait & Special & 
    \textbf{Format:} \texttt{[time]}\newline
    \textbf{Details:} Time to wait in seconds. \\
    \addlinespace
    
    done & Special & 
    \textbf{Format:} \texttt{[]}\newline
    \textbf{Details:} Signals successful completion of the entire task. \\
    \addlinespace
    
    fail & Special & 
    \textbf{Format:} \texttt{[]}\newline
    \textbf{Details:} Signals that the task is impossible to complete. \\
    
    \bottomrule
  \end{tabularx}

  \captionsetup{skip=3pt, position=bottom}
  \caption{Action Space of \name.}
  \label{tab:action_specifications}
  
\end{table*}

\subsection{Action Space}
\label{sec:action_space}

Designing an optimal action space is of paramount importance for CUAs, and its impact on task success rates and execution efficiency may even outweigh that of the system architecture itself. Building upon the insights and foundational methodologies of the AgentS series work~\citep{gonzalez2025unreasonable}, we have formulated our design based on the following core principles:

\begin{itemize}[leftmargin=*]
    \item \textbf{Cross-Platform Abstraction.} For cross platform CUAs, it is essential to establish a set of intermediate-level actions, which are subsequently mapped to platform-specific executable primitives (always be \texttt{PyAutoGUI}\footnote{\url{https://pyautogui.readthedocs.io/}}).
    \item \textbf{Conciseness and Generality.} For general-purpose CUAs, the action space must be concise, pragmatic, and universally applicable. Redundant actions inevitably introduce additional context overhead and cognitive load, whereas overly specialized actions, such as manually constructed MCP tools, impose severe constraints on the agent's operational versatility.
\end{itemize}

Consequently, our action space is detailed in Tab.~\ref{tab:action_specifications}.
We categorize the action space into three distinct types: GUI actions, proprietary actions, and special actions. Notably, we have excluded application-specific primitives, such as \texttt{set\_cell\_value} for LibreOffice Calc, from our action space. While precise table localization remains a challenge for current Grounders, our extensive testing suggests that \texttt{call\_code\_agent} serves as a superior alternative for such granular tasks. Extending beyond specific benchmarks, our design of action space and overall framework enables high performance without relying on defining specialized actions for every software feature. Nevertheless, the design of optimal action spaces, including the action design and the action parameters' design, remains an open question worthy of further investigation.

\section{More Results}
\label{sec:extra_exp}

In this section, we first show the more results on OSWorld, then present the comprehensive evaluation results for both WindowsAgentArena and MacOSArena. We also explored the instruction rewriting method followed by an in-depth analysis of various statistical metrics.

\subsection{More Results on OSWorld}

\begin{figure}[H]
    \centering
    \includegraphics[width=\columnwidth]{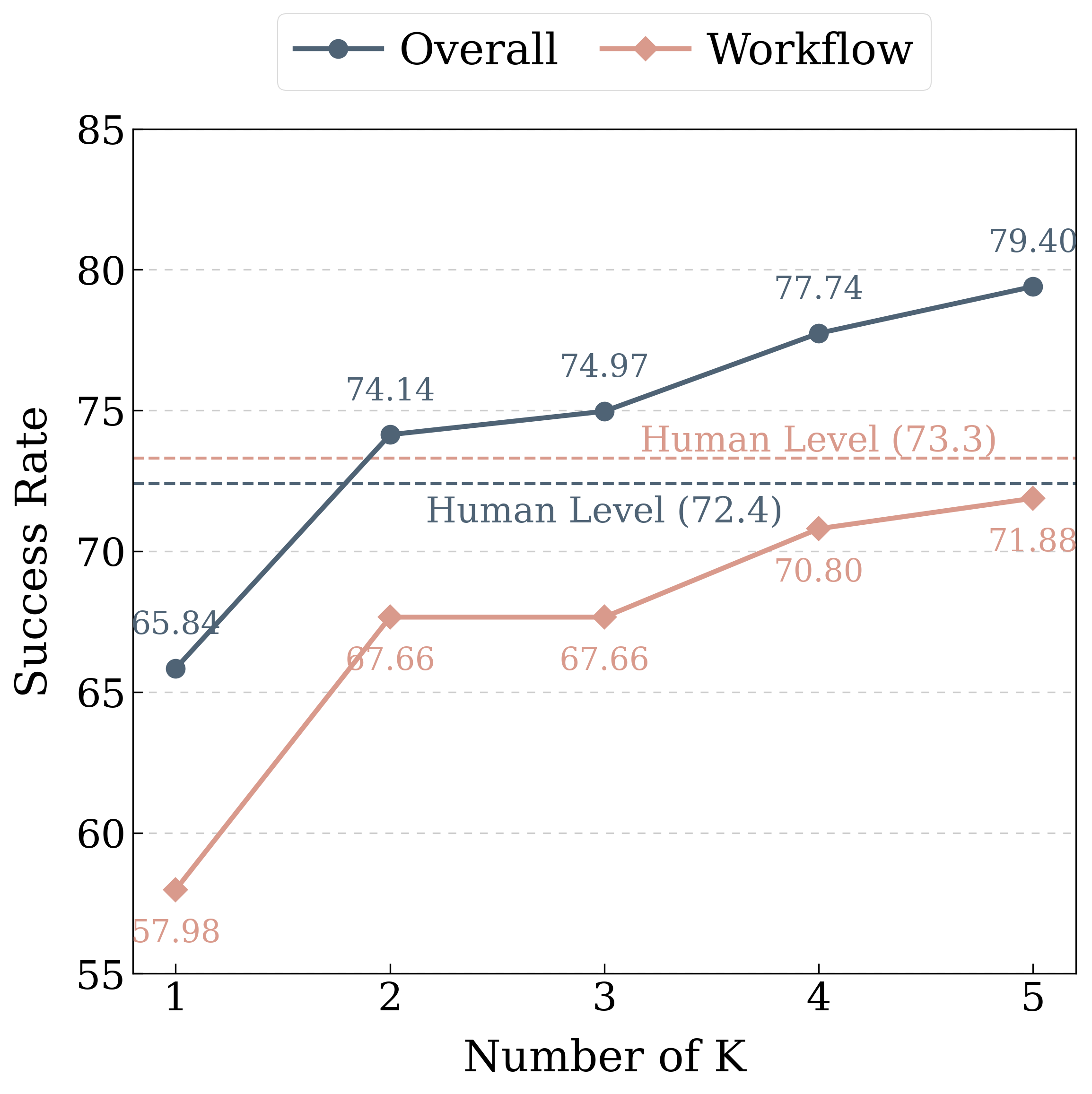} 
    \captionsetup{skip=0pt, position=bottom}
    \caption{The Pass@K results on OSWorld. All experiments are carried out with GPT-5 and 100 steps limit.}
    \label{fig:pass_k}
\end{figure}
\paragraph{Results with Pass@K.} As shown in Fig.~\ref{fig:pass_k}, \name\ surpasses human performance~(72.4\%) at Pass@2~(74.14\%) and achieves a success rate approaching 80\% at Pass@5~(79.4\%) on OSWorld. To explore the performance limits, we incrementally increased the temperature of both the Orchestrator and Reflection-Memory Agent by 0.1 at each pass, reaching a 0.5 increase at Pass@5 to encourage diverse solution attempts. This demonstrates, first, that our framework has a high performance ceiling and can attain excellent results through such test-time scaling. Second, consistent with our analysis, zero-shot agentic frameworks exhibit substantial stochasticity on GUI tasks, which is evident in both our experimental results and qualitative observations. Future work should therefore prioritize improving deployment-time stability. 

\begin{table}[t]
  \centering
  \small
  
  \begin{tabularx}{\linewidth}{l c c}
    \toprule
    \textbf{Method} & \textbf{Avg.}(\%) \\
    
    \midrule
    Qwen3-VL-8B-Instruct~\citeyearpar{bai2025qwen3} & 33.9 \\
    Qwen3-VL-8B-Thinking~\citeyearpar{bai2025qwen3} & 33.9 \\
    Qwen3-VL-32B-Instruct~\citeyearpar{bai2025qwen3} & 32.6 \\
    Qwen3-VL-32B-Thinking~\citeyearpar{bai2025qwen3} & 41.0 \\
    
    \midrule

    \name \ w/ Qwen3-VL-8B-I. & 33.9 ($\uparrow0\%$) \\
    \name \ w/ Qwen3-VL-8B-T. & 39.1 ($\uparrow15.3\%$) \\
    \name \ w/ Qwen3-VL-32B-I. & 46.9 ($\uparrow43.9\%$) \\
    \name \ w/ Qwen3-VL-32B-T. & 50.2 ($\uparrow22.4\%$) \\
    
    \bottomrule
  \end{tabularx}
  \captionsetup{skip=3pt, position=bottom}
  \caption{Impact of reasoning proficiency in OSWorld with Qwen3-VL series+UI-TARS-1.5-7B and 50 steps limit. Values in parentheses indicate the relative improvement over the corresponding base models.}
  \label{tab:appendix_qwen_thinking}
  \vspace{-0.5cm}
\end{table}
\paragraph{Impact of Thinking.}

We employed the Qwen3-VL family (spanning 8B and 32B scales, with both Instruct and Thinking variants) as based VLMs to investigate the impact of reasoning capabilities on OSWorld performance. As shown in Tab.~\ref{tab:appendix_qwen_thinking}, performance generally correlates positively with model scale and reasoning proficiency. Notably, while the vanilla 8B baselines exhibit identical performance, their integration with \name\ revealed a significant divergence: the Thinking variant surpassed its Instruct counterpart by approximately 5\%. This disparity suggests that our framework's decoupling of reasoning and localization effectively alleviates the cognitive load of multi-tasking, a benefit that is particularly pronounced when the based VLM possesses latent reasoning strengths. Furthermore, although the advantage of Thinking models extends to the 32B series, the 32B-Instruct model demonstrated high relative gains. However, given that the vanilla 32B-Instruct baseline underperforms even the 8B-Instruct baseline, we attribute this irregularity to the baseline's instability rather than a structural advantage. Conclusively, Thinking models prove to be the optimal based VLMs, as our framework heavily relies on and effectively amplifies the strong intrinsic reasoning capabilities.

\paragraph{Impact of Instruction Rewriting.}

Given the colloquial ambiguity inherent in current benchmark's user instructions and their frequent misalignment with the initial visual state (\eg, omitting specific target applications), we initially explored an instruction rewriting mechanism to mitigate task deviation. Specifically, we employed a VLM that accepts the raw user instruction and the initial screenshot to generate a refined instruction via a predefined prompt. The purpose is to produce professional, concise instructions that are visually grounded. For instance:
\begin{itemize}[leftmargin=*]
    \item \textbf{Original:} I need to include the experiment results from \path{~/Documents/awesome-desktop/expe-results.xlsx} into the currently writing report. Specifically, extract the results of GPT-4 and insert a table into the ``Main Results'' section of my report.
    \item \textbf{Rewritten:} Insert a table into the ``Main Results'' section of the open document \path{awe_desk_env.docx} in LibreOffice Writer containing the GPT-4 experiment results extracted from \path{~/Documents/awesome-desktop/expe-results.xlsx}.
\end{itemize}

\begin{table}[H]
    \centering

    \begin{tabular}{lc}
        \toprule
        \textbf{Method} & \textbf{Workflow} \\
        \midrule
        w/ instruction rewriting & 51.09 \\
        w/o instruction rewriting (Ours) & 54.86 \\
        \bottomrule
    \end{tabular}
    
    \captionsetup{skip=3pt, position=bottom}
    \caption{Impact of insrtuction rewriting in OSWorld with GPT-5+UI-TARS-1.5-7B and 50 steps limit.}
    \label{tab:rewrite_ablation}
    \vspace{-0.3cm}
    
\end{table}

However, as shown in Tab.~\ref{tab:rewrite_ablation}, preliminary experiments indicated that differences in performance were negligible. Furthermore, we determined that altering the test instructions might compromise the integrity of the evaluation (i.e., potential data leakage or simplification). Consequently, we pivoted to a keeping first image strategy. Nevertheless, we maintain that instruction rewriting remains an indispensable component for robust real-world deployment.

\begin{table}[H]
    \centering

    \begin{tabular}{lc}
        \toprule
        \textbf{Method} & \textbf{Avg.} \\
        \midrule
        w/o Coder & 57.42 \\
        w/ Coder(Ours) & 63.61 \\
        \bottomrule
    \end{tabular}

    \captionsetup{skip=3pt, position=bottom}
    \caption{Impact of Coder in OSWorld with GPT-5+UI-TARS-1.5-7B and 50 steps limit.}
    \label{tab:coder_ablation}
    \vspace{-0.3cm}
    
\end{table}

\paragraph{Impact of Coder.}

While the Coder is not the central contribution of our framework, we performed an ablation study to assess its individual contribution, as detailed in Tab.~\ref{tab:coder_ablation}. For the ``w/o Coder'' setting, we disabled the \texttt{call\_code\_agent} action and eliminated all relevant prompts, constraining the model to rely solely on GUI interactions. The results reveal a performance degradation of approximately 6.2\% in the absence of the Coder. This finding not only underscores the robustness of the Coder in handling tasks such as batch file processing and content editing, but also substantiates the necessity of a hybrid GUI-API paradigm for the advancement of CUAs.

\begin{figure}[H]
    \centering
    \includegraphics[width=\columnwidth]{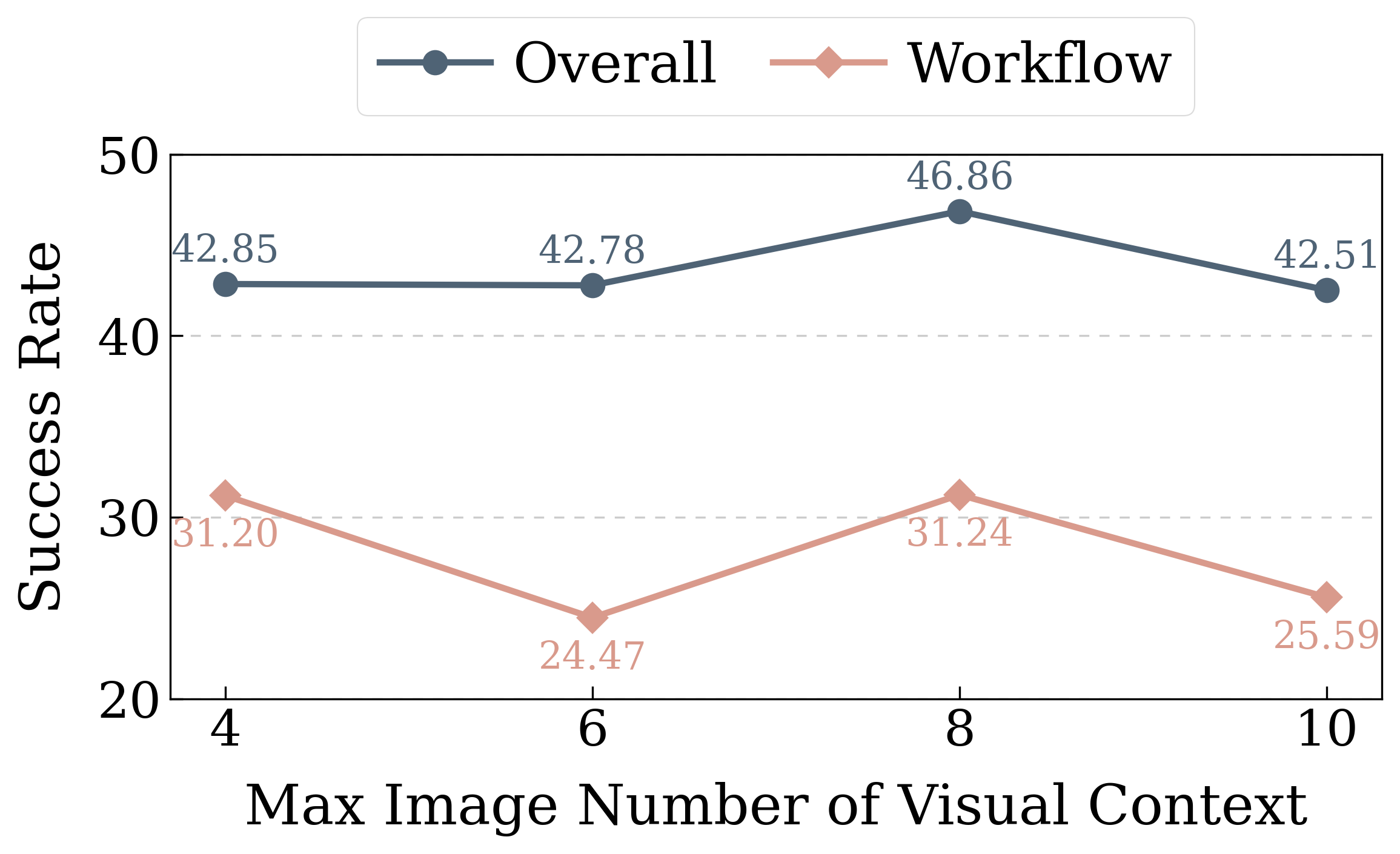} 
    \captionsetup{skip=0pt, position=bottom}
    \caption{Discussion on the impact of varying the maximum number of images in agent’s trajectory. All experiments are carried out with Qwen3-VL-32B-Instruct and UI-TARS-1.5-7B.}
    
    \label{fig:different_max_images}
    \vspace{-0.3cm}
\end{figure}

\paragraph{Impact of Visual Context Length.}

In this subsection, we investigate the sensitivity of performance to the maximum visual context length. As illustrated in Fig.~\ref{fig:different_max_images}, the quantity of retained images significantly influences framework efficacy. A sparsity of images fails to provide adequate context regarding interaction history, whereas an excessive visual load saturates the context window, degrading the model's reasoning capabilities due to information overload. This aligns with our earlier insights regarding the context-reasoning trade-off. Empirically, our results confirm that a cap of 8 images strikes the optimal balance, achieving a peak performance of 46.86\%.

\begin{table*}[!t]
  \centering
  \small

  \resizebox{\textwidth}{!}{
  \begin{tabularx}{\textwidth}{ l c *{8}{>{\centering\arraybackslash}X} }
    \toprule
    \multirow{2}{*}{\textbf{Method}} & \multirow{2}{*}{\textbf{Step}} 
    & \multicolumn{8}{c}{\textbf{Success Rate}(\%)}  \\ 
    \cmidrule(lr){3-10} 
    & & \textbf{Office} & \textbf{Web} & \textbf{Sys.} & \textbf{Code} & \textbf{Media} & \textbf{Util.} & \textbf{Inf.} & \textbf{Avg.} \\
    
    \midrule

    Qwen3-VL-32B-Instruct$^{\clubsuit}$ & 50 & 19.05 & 49.66 & 54.17 & 21.05 & 42.19 & 25.00 & 0.00 & 31.68 \\
    UI-TARS-1.5-7B & 50 & - & - & - & - & - & - & - & 42.10 \\
    UI-TARS-2 & 50 & - & - & - & - & - & - & - & 50.60 \\
    Agent S3 w/ GPT-5 & 50 & - & - & - & - & - & - & - & 54.10 \\
    Agent S3 w/ GPT-5 & 100 & - & - & - & - & - & - & - & 56.60 \\
    
    \midrule
    
    \name \ w/ Qwen3-VL-32B-Inst. & 50 & 26.19 & 46.33 & \underline{75.00} & 47.37 & 27.90 & 41.67 & \textbf{69.23} & 45.32 \\
    
    \name \ w/ GPT-5-Mini & 50 & \underline{42.86} & \textbf{73.00} & \textbf{79.17} & \textbf{68.42} & \underline{48.66} & \underline{66.67} & \textbf{69.23} & \underline{62.15}  \\
    \name \ w/ GPT-5 & 50 & \textbf{54.76} & \textbf{73.00} & \underline{75.00} & 42.11 & \textbf{70.09} & \textbf{75.00} & 53.85 & \textbf{63.45} \\ 
    
    \bottomrule
  \end{tabularx}
  }
  \captionsetup{skip=3pt, position=bottom}
  \caption{Main results of \name\ on WindowsAgentArena which has 154 tasks. Office includes LibreOffice Writer and LibreOffice Calc tasks; Web (Web Browsing) includes Edge and Chrome tasks; Sys.(Windows System) includes File Explorer and Settings tasks; Code includes VSCode tasks; Media(Media \& Video) includes VLC tasks; Util.(Windows Utilities) includes Notepad, Clock, Paint and WindowsCalc tasks; Inf.(Infeasible) includes 13 infeasible tasks. $^{\clubsuit}$ represents the result reproduced by us, and the others are sourced from the original papers.}
  \label{tab:appendix_more_results_waa}
\end{table*}
\begin{table*}[htbp]
  \centering
  \small

  \begin{tabularx}{\textwidth}{ l c *{3}{>{\centering\arraybackslash}X} }
    \toprule
    \multirow{2}{*}{\textbf{Method}} & \multirow{2}{*}{\textbf{Step}} 
    & \multicolumn{3}{c}{\textbf{Success Rate}(\%)}  \\
    \cmidrule(lr){3-5}
    & & \textbf{Single-Apps} & \textbf{Multi-Apps} & \textbf{Avg.} \\
    \midrule
    
    GPT-4o~\citeyearpar{hurst2024gpt} & 50 & 3.57 & 0.00 & 1.59 \\
    Claude-3.7-Sonnet~\citeyearpar{anthropic2025claude37} & 50 & 14.29 & 2.86 & 7.94 \\
    Aguvis-72B~\citeyearpar{xu2024aguvis} & 50 & 0.00 & 0.00 & 0.00 \\
    UI-TARS-1.5-7B~\citeyearpar{qin2025uitars} & 50 & 14.29 & 2.86 & 7.94 \\
    UI-TARS-72B-DPO~\citeyearpar{qin2025uitars} & 50 & 14.29 & 5.71 & 9.52 \\
    Qwen2.5-VL-72B~\citeyearpar{bai2025qwen25vl} & 50 & 7.14 & 0.00 & 3.17 \\
    Qwen3-VL-32B-Inst.~\citeyearpar{bai2025qwen3} & 50 & 17.86 & 0.00 & 7.94 \\
    
    \midrule
    
    \name \ w/ Qwen3-VL-32B-Inst. & 50 & \underline{32.14} & \underline{8.57} & \underline{19.05} \\
    \name \ w/ GPT-5-Mini & 50 & \textbf{57.14} & \textbf{37.14} & \textbf{46.03} \\

    \bottomrule
  \end{tabularx}

  \captionsetup{skip=3pt, position=bottom}
  \caption{Main results of \name\ on MacOSArena which has 63 tasks. Single-Apps includes Calendar, Clock, Finder, Mac System Settings, Notes, Reminders, Safari, Terminal, and VSCode tasks; Multi-Apps includes combined tasks of two domains.}
  \label{tab:appendix_more_results_mac}

  \vspace{-0.3cm}
\end{table*}

\subsection{More Results on WindowsAgentArena \& MacOSArena}

\label{sec:extra_results_win_mac}
\paragraph{WindowsAgentArena.} The comprehensive results on WindowsAgentArena are presented in Tab.~\ref{tab:appendix_more_results_waa}. Our analysis yields the following insights: First, the three based VLMs configurations of \name\ demonstrate progressively higher performance, which aligns with general expectations. In the particularly challenging Office domain, our framework achieves a top score of 54.76\%. Next, the tasks in WindowsAgentArena are partially inherited from OSWorld, while the remainder are system-level software tasks. This benchmark’s lack of complex multi-apps tasks may mean that our framework has not yet fully exploited its potential, despite already achieving SOTA performance, indicating room for further gains. 
Finally, our case study indicates that adapting to the specific characteristics of the Windows environment remains a challenge for general-purpose VLMs (\eg, Qwen3-VL series), but our framework addresses this adaptation bottleneck. Taking Qwen3-VL-32B-Instruct as a representative example, our method effectively enhances capability across nearly every domain, achieving a relative increase of approximately 24.6\% on average compared to the vanilla baseline. This result reinforces our belief that \name\ provides substantial benefits, regardless of the based VLM's scale or initial strength.

\paragraph{MacOSArena.} The comprehensive results on MacOSArena are presented in Tab.~\ref{tab:appendix_more_results_mac}. Due to the cost constraint, we didn't use GPT-5 as the based VLMs. Our analysis yields the following insights:
First, \name\ establishes a new SOTA. Even with the Qwen3-VL-32B-Instruct, \name\ achieves 19.05\%, surpassing all competing baselines. Furthermore, \name\ with GPT-5-Mini reaches a remarkable 46.03\%, significantly outperforming other methods, suggesting that GPT-5-Mini is a highly cost-effective and capable choice for MacOS tasks. In stark contrast, existing strong baselines struggle severely on this benchmark, with some yielding 0\% success rates. This highlights a critical generalization gap: models trained primarily on Linux or Windows fail to adapt to the MacOS environment. We attribute this to two factors: (1) Data scarcity for the MacOS domain~\cite{liu2025scalecua}; and (2) Intrinsic UI challenges, such as the minute ``traffic light'' window controls (red/yellow/green buttons) which are difficult for models to locate, complicating window management and app switching.
Besides, despite these environmental challenges, \name\ consistently increases model capabilities across scales, notably boosting Qwen3-VL-32B-Instruct by approximately 140\% relative to its vanilla baseline. We emphasize that future CUA research must not overlook the MacOS platform. True cross-platform generality requires rigorous training and testing on MacOS environment to bridge the current performance gap.

\begin{figure*}[htbp]
    \centering
    \begin{subfigure}[b]{0.48\linewidth}
        \centering
        \includegraphics[width=\linewidth]{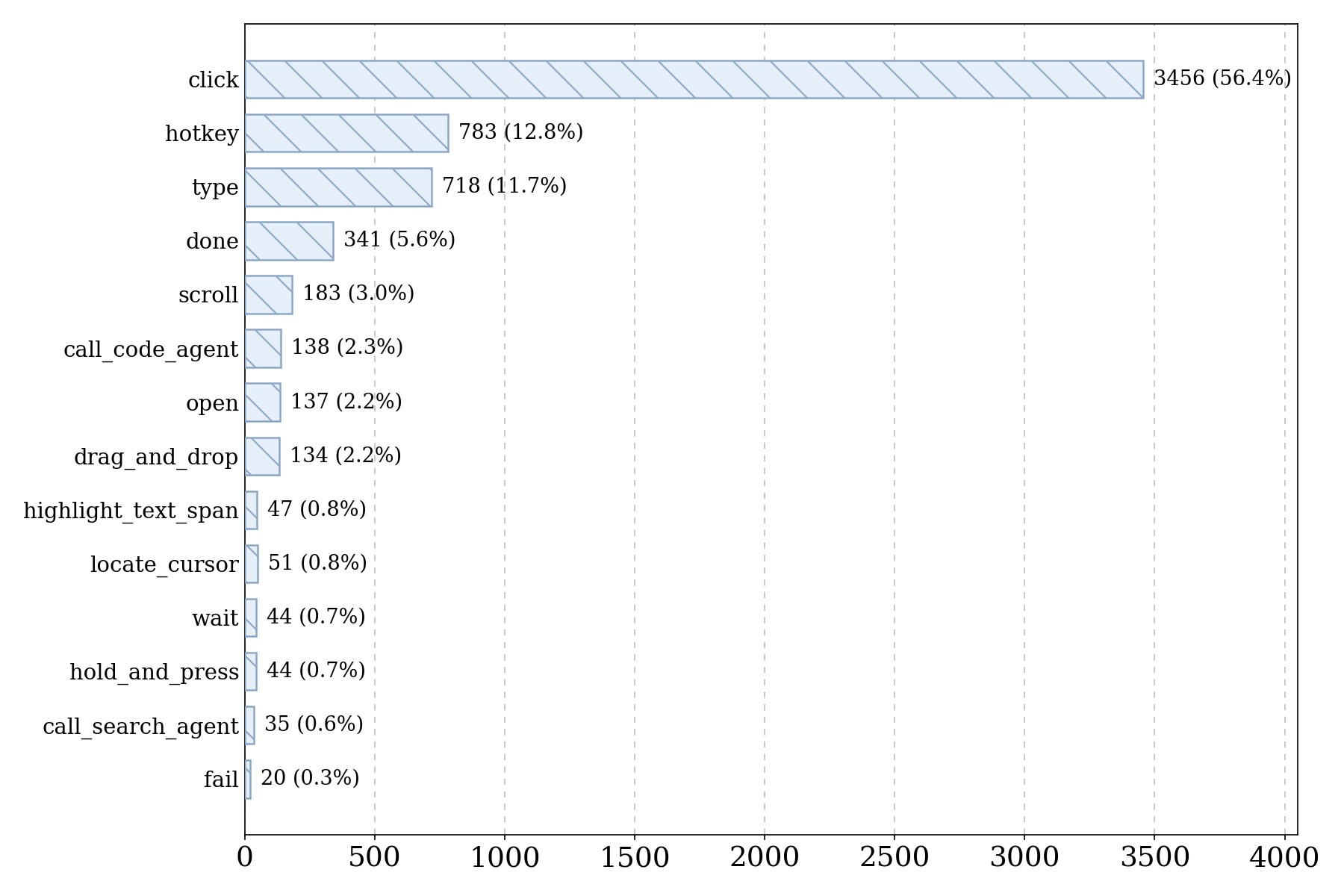}
        \caption{Different actions and their invocation counts}
        \label{fig:other_stats_a}
    \end{subfigure}
    \hfill
    \begin{subfigure}[b]{0.48\linewidth}
        \centering
        \includegraphics[width=\linewidth]{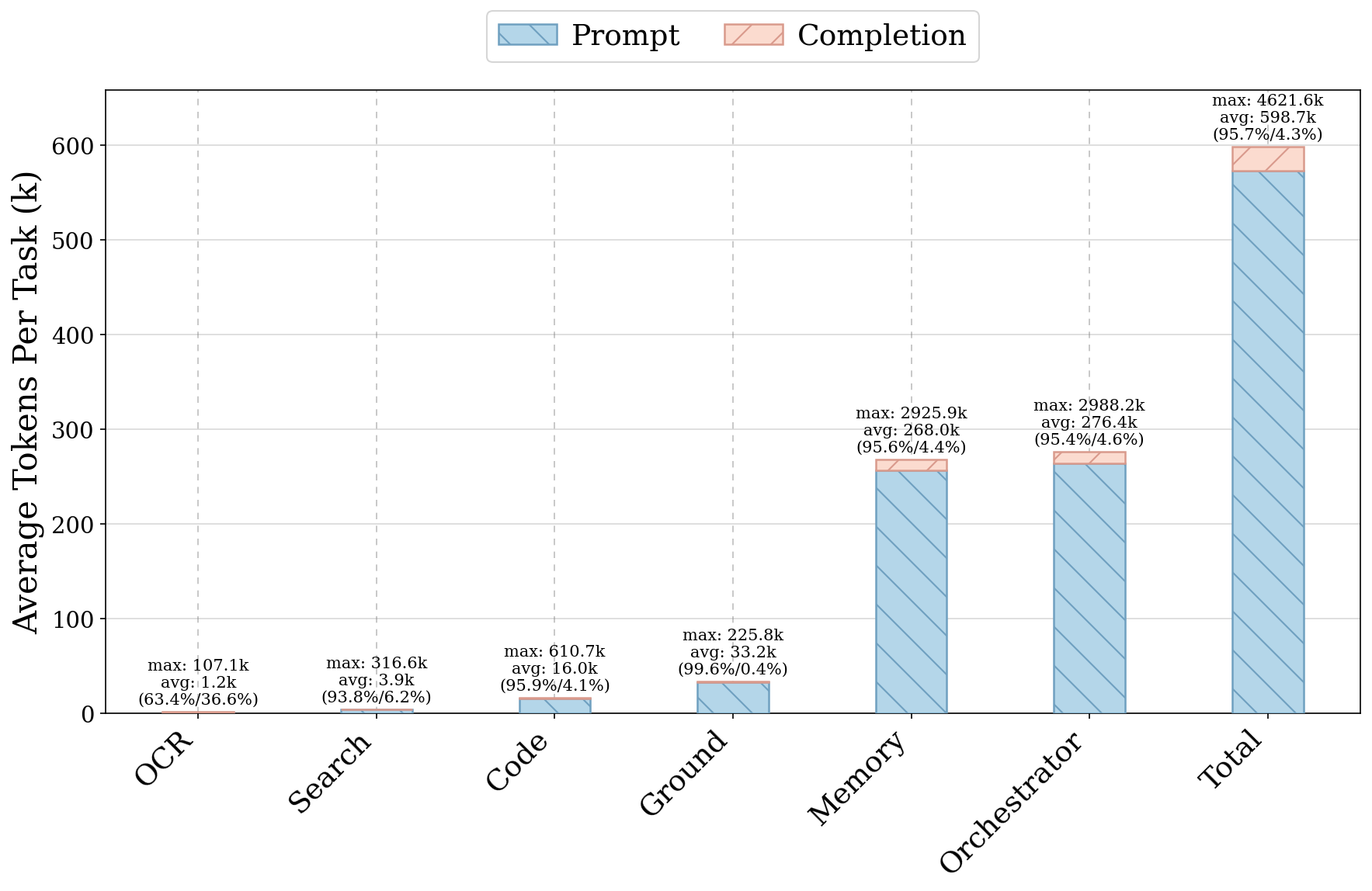}
        \caption{Average token usage on each task for each agents. }
        \label{fig:other_stats_b}
    \end{subfigure}
    
    \begin{subfigure}[b]{0.48\linewidth}
        \centering
        \includegraphics[width=\linewidth]{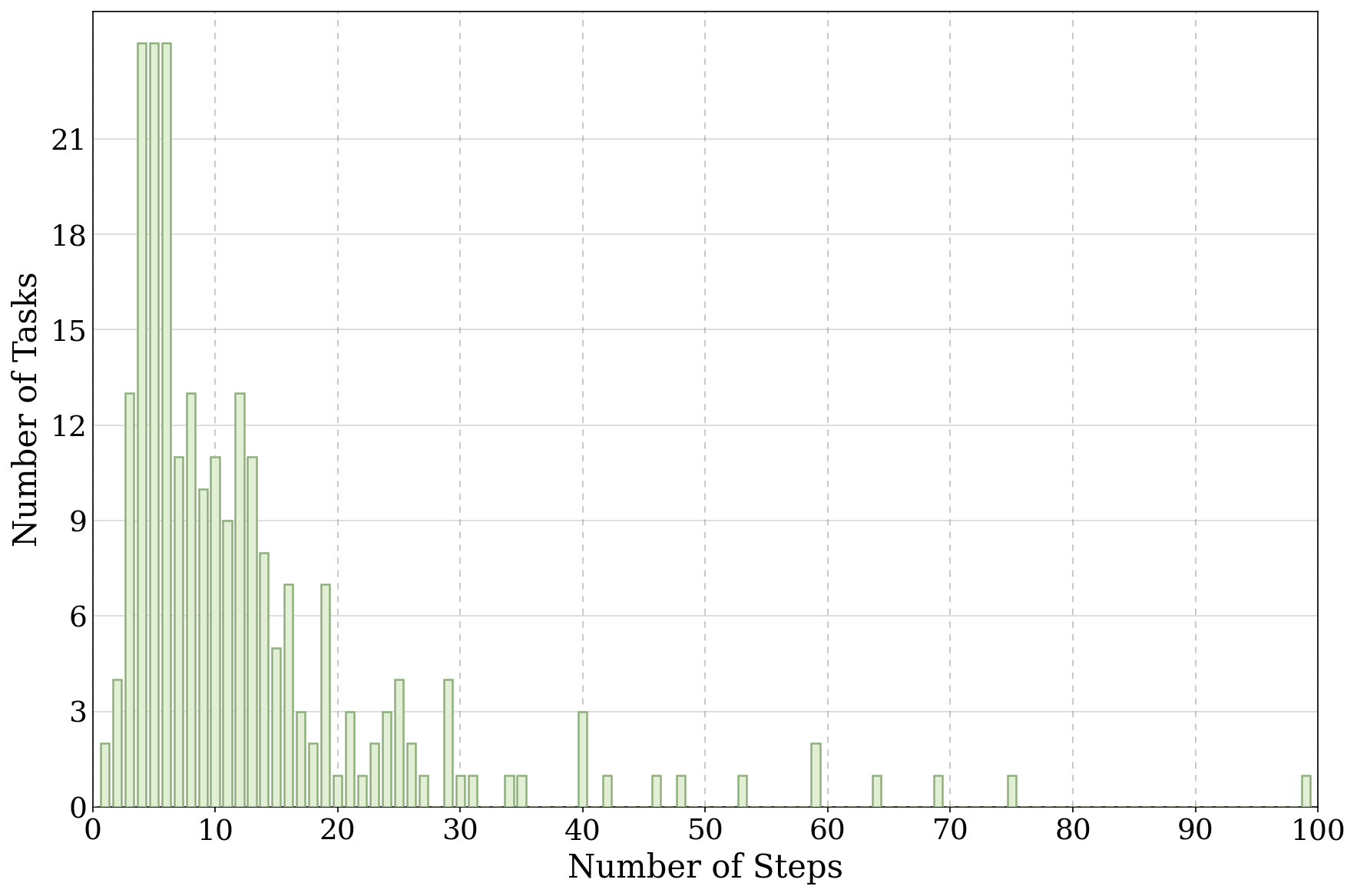}
        \caption{Distribution of steps for successful tasks}
        \label{fig:other_stats_c}
    \end{subfigure}
    \hfill
    \begin{subfigure}[b]{0.48\linewidth}
        \centering
        \includegraphics[width=\linewidth]{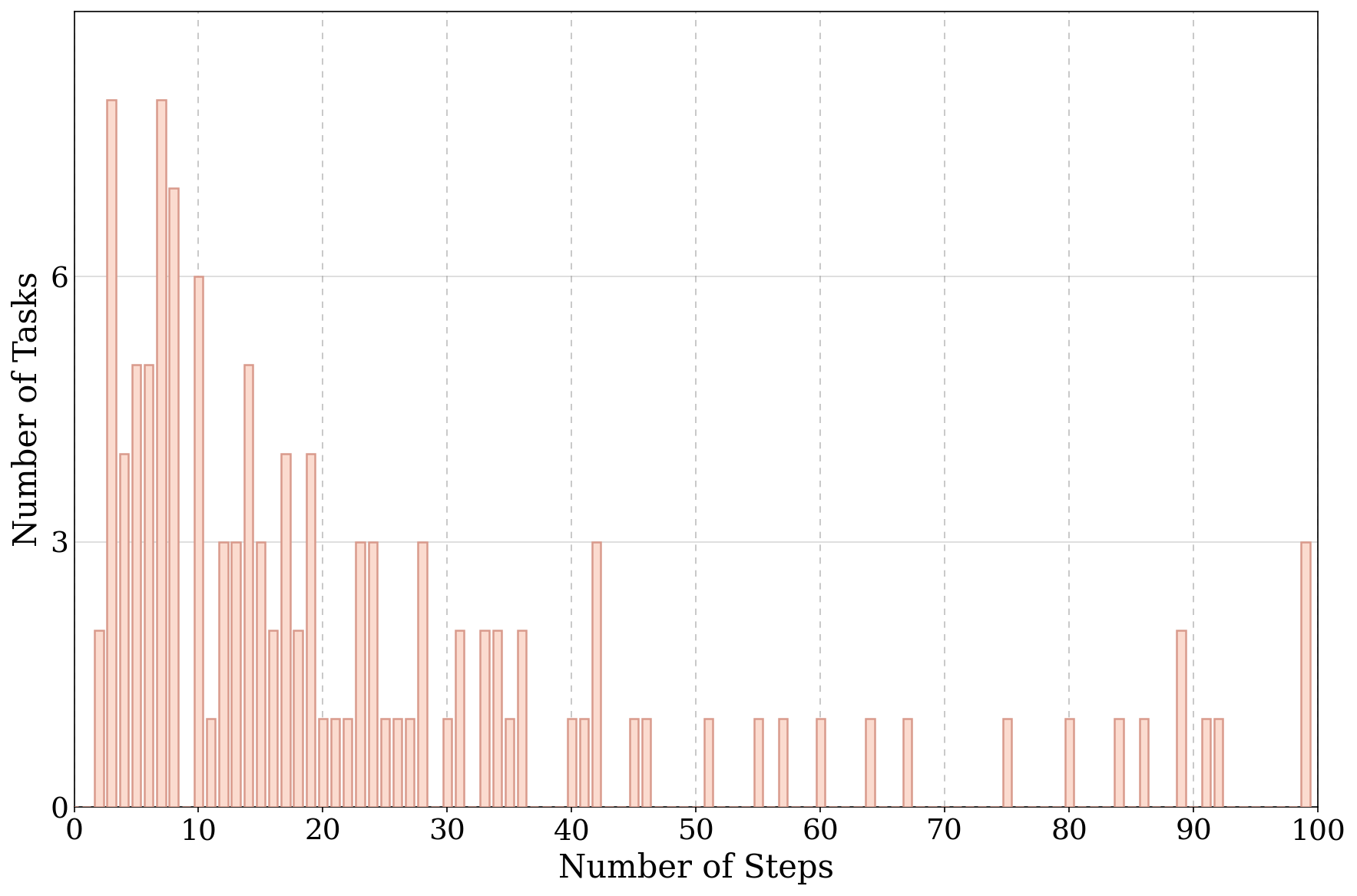}
        \caption{Distribution of steps for unsuccessful tasks}
        \label{fig:other_stats_d}
    \end{subfigure}
    
    \caption{More statistics on OSWorld. All experiments are carried out with GPT5 and 100 steps limit.}
    \captionsetup{skip=-4pt, position=bottom}
    \label{fig:other_stats}
\end{figure*}

\subsection{Other Statistics}

We conducted a comprehensive visual analysis of token usage, action utilization, and the distribution of steps for both successful and failed tasks within the OSWorld evaluation, as illustrated in Fig.~\ref{fig:other_stats}. 

Regarding the step distribution, we observed that both successful and failed tasks are predominantly concentrated within the first 15 steps. Given that standard testing protocols typically allow for 50 or 100 steps, this reveals two critical insights. 
First, \name \ fail to utilize the maximum available steps to achieve a theoretical optimum, often exhibiting a tendency to prematurely terminate tasks with high confidence. This suggests a need to incentivize agents to utilize remaining steps for verification after initial task completion. 
Second, this highlights a design flaw regarding infeasible tasks in benchmarks. We concur with prior analyses~\citep{liu2025pc} that the evaluation of infeasible tasks is susceptible to gaming; aside from obvious factual contradictions (\eg, ``install Python 4''), most infeasible tasks require extensive exploration to confirm unreachable status. Most CUAs default to outputting \texttt{fail} only upon reaching the step limit, thereby easily get these scores. In our experiments, one task succeeded in this manner at 100 steps, introducing unfairness and necessitating a re-evaluation of how infeasible tasks are scored. 

In terms of action utilization, \texttt{click} is indisputably the most frequent operation, accounting for 56.4\% of actions. Together with \texttt{hotkey}, \texttt{type} and \texttt{scroll}, these four fundamental GUI interactions comprise 83.9\% of the total usage. Notably, while the \texttt{call\_search\_agent} action was invoked only 35 times (0.6\%), our manual verification confirmed that 85\% of these searches provided valuable tutorials, demonstrating the efficacy of the \textit{Visual-Centric Search as a Tool} paradigm where the agent invokes external help only when strictly necessary. 

Finally, the token usage analysis indicates that the RMA and Orchestrator are the most resource-intensive modules, averaging approximately 270k tokens per task, with roughly 96\% attributed to context prompts and 4\% to completion. The widely used General Grounder follows with an average of 33.2k tokens per task, while specialized modules like the Coder, Searcher, and OCR Grounder exhibit significantly lower average usage as they are not invoked for every task.

Furthermore, we conducted a statistical analysis on the trigger types of the our message protocol, which serves as the communication bridge between the RMA and the Orchestrator within a complete experimental run, as presented in Tab.~\ref{tab:reflection_message_protocol}. As observed, approximately 90\% of the GUI errors detected by the step summary module were identified as \textit{GUI Error} by the RMA and fed back to the Orchestrator. Meanwhile, roughly 50\% of the loops identified by our loop detection algorithm were classified by the RMA as \textit{Lack of Tutorial}. Regarding the distribution of the message protocol feedback types, approximately 75\% of steps were judged as normal operations~(on track), whereas a significant portion—up to 25\%—were flagged as various types of errors~(off track). This distribution not only demonstrates the rigorous nature of our message protocol but also highlights the indispensability of the reflection mechanism given the current limitations of VLMs capabilities. To further investigate the alignment between the error types output by the RMA and the ground truth, we manually selected 100 steps where the RMA output was \textit{GUI Error} and invited human experts to analyze the actual status of these steps. The results indicate that in approximately 90 cases, actual GUI errors occurred, with only about 10 cases attributable to RMA hallucinations~(refer to Sec.~\ref{sec:discussion}). This finding further underscores the effectiveness of our RMA's reflection checks.

\begin{table*}[tbp]
    \centering
    
    \begin{tabular}{lccccc}
        \toprule
        \multirow{2}{*}{\textbf{Aux. Det.}} & \multicolumn{5}{c}{\textbf{Message Protocol Statistics}} \\
        \cmidrule{2-6}
         & \textbf{GUI Error} & \textbf{Lack of Tutorial} & \textbf{Code Error} & \textbf{Other Error} & \textbf{Normal} \\
        \midrule
        \textbf{GUI Error} & 1133 (18.5\%) & 52 (0.8\%) & 2 (0.0\%) & 9 (0.1\%) & 49 (0.8\%) \\
        \textbf{Loop Error} & 1 (0.0\%) & 29 (0.5\%) & 0 (0.0\%) & 0 (0.0\%) & 29 (0.5\%) \\
        \textbf{Normal} & 30 (0.5\%) & 143 (2.3\%) & 18 (0.3\%) & 36 (0.6\%) & 4600 (75.0\%) \\
        \bottomrule
    \end{tabular}

    \captionsetup{skip=3pt, position=bottom}
    \caption{Message Protocol statistics on OSWorld with GPT-5+UI-TARS-1.5-7B and 100 steps limit.}
    \label{tab:reflection_message_protocol}
    \vspace{-0.3cm}
\end{table*}

\section{Case Study}
\label{sec:case_study}

In this section, to better demonstrate the strengths and limitations of our framework, we conduct a qualitative analysis of specific success and failure cases observed during experiments.

\subsection{Correct Case}
\label{sec:case_study_correct}

\begin{figure*}[htbp]
    \centering
    \includegraphics[width=\linewidth]{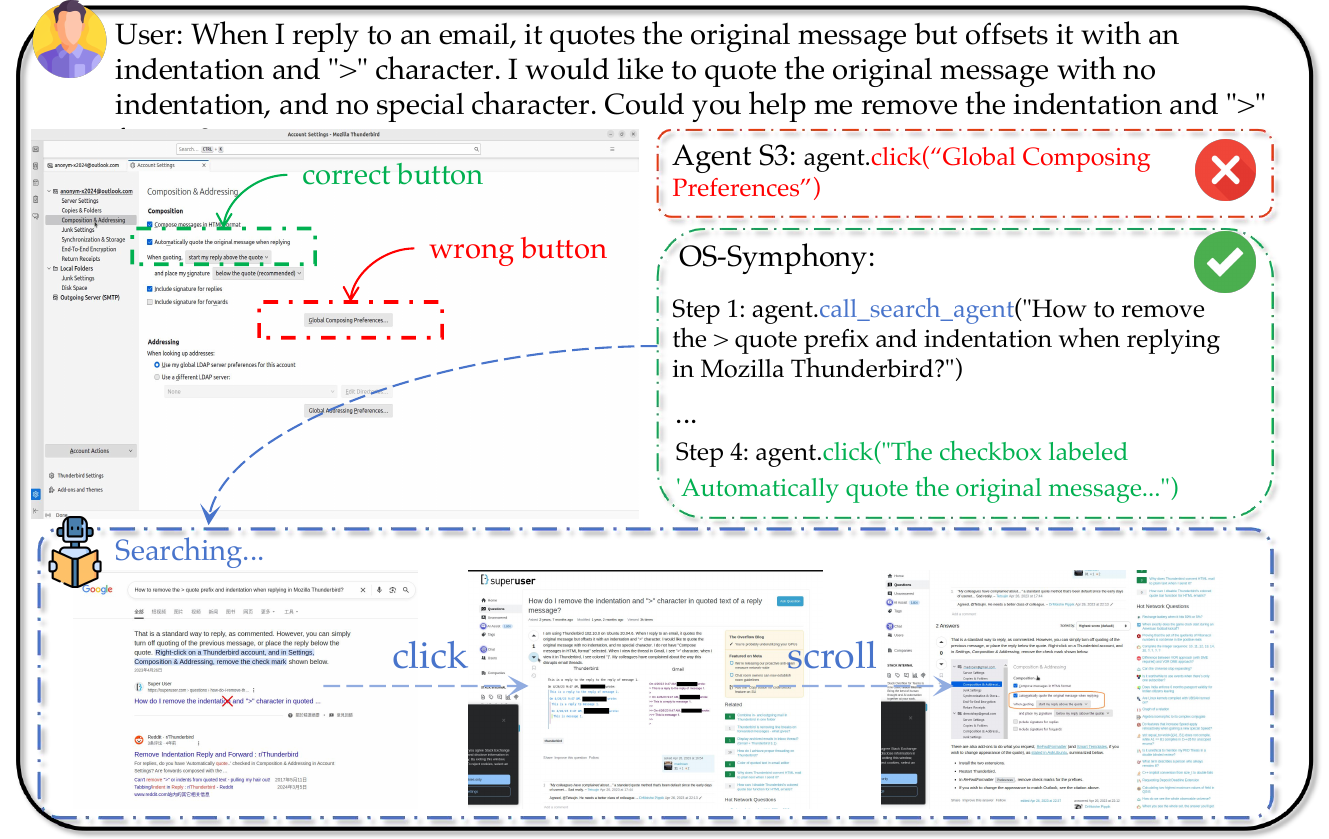} 
    
    \captionsetup{skip=0pt, position=bottom}
    \caption{A successful case of \name benefiting from the Multimodal Searcher.}
    \label{fig:correct_case_search}
\end{figure*}

\noindent \textbf{Effectiveness of Multimodal Searcher.}
\name\ incorporates a Searcher designed to mimic human search behavior. Fig.~\ref{fig:correct_case_search} illustrates a successful instance on OSWorld where the model is tasked with utilizing a built-in feature of Thunderbird. After navigating to the correct page, the primary baseline, Agent S3, suffered from a lack of domain knowledge. It clicked an incorrect button, which visually resembled a settings option but was functionally irrelevant, and subsequently became trapped in an erroneous loop until the maximum step limit was exhausted. In contrast, \name\ invoked the Searcher at the first step. Through Google Chrome and a series of GUI actions, the Searcher navigated to the ``Superuser" website and synthesized a relevant tutorial. Guided by this external knowledge, our framework correctly identified and clicked the target button at Step 4, successfully completing the task.

\begin{figure*}[htbp]
    \centering
    \includegraphics[width=\linewidth]{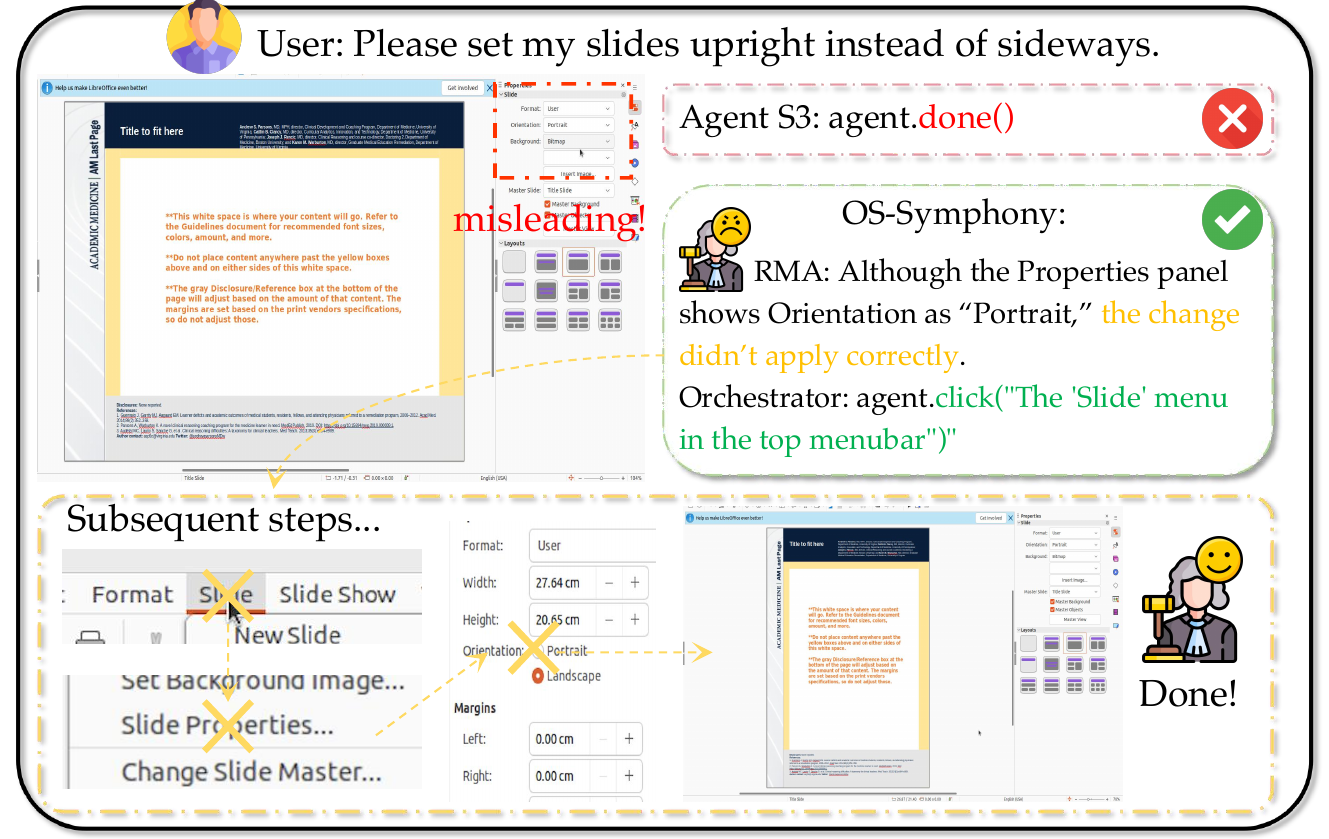} 
    
    \captionsetup{skip=0pt, position=bottom}
    \caption{A successful case of \name benefiting from the RMA.}
    \label{fig:correct_case_rma}
\end{figure*}

\noindent \textbf{Effectiveness of Reflection-Memory Agent.} \name\ features a refined reflection mechanism where the RMA verifies actions based on history summaries and the screenshot of the current step. As illustrated in Fig.~\ref{fig:correct_case_rma}, in a task requiring a change in slide orientation, both \name\ and Agent S3 initially attempted to modify the setting via the ``Properties'' panel. However, the screen remained unchanged following this operation. Agent S3 erroneously concluded that the modification was successful and prematurely terminated the task with a \texttt{done} output. In contrast, our framework received feedback from the RMA indicating that the visual state remained unaltered and the action had failed. Consequently, the Orchestrator pivoted to an alternative strategy to complete the task. This demonstrates the benefits arising from the collaboration between the RMA and the Orchestrator, validating the effectiveness of our framework's design.

\begin{figure*}[htbp]
    \centering
    \includegraphics[width=\linewidth]{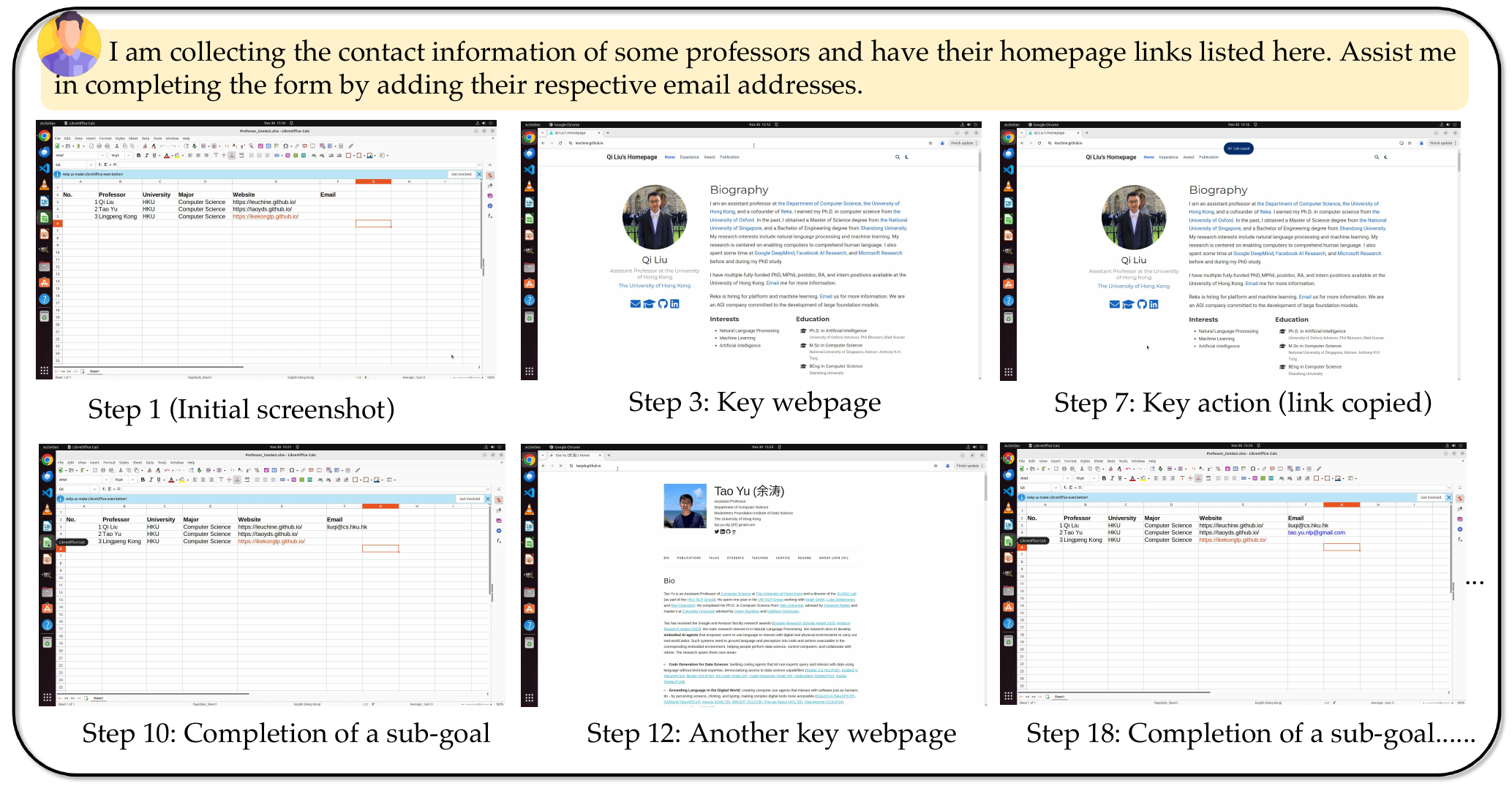} 
    
    \captionsetup{skip=0pt, position=bottom}
    \caption{An example to show the milestone identification mechanism of \name.}
    \label{fig:correct_case_milestones}
\end{figure*}

\noindent \textbf{Milestone Identification.} To conserve the agent's context window, the RMA selectively saves screenshots exclusively from ``milestone'' steps. Fig.~\ref{fig:correct_case_milestones} illustrates our criteria for selecting these images. First, the initial screenshot is invariably classified as a milestone. Since textual instructions often lack explicit references to specific webpages or files, the initial visual state complements the text to fully define the task requirements. Additionally, pivotal actions are designated as milestones, such as navigating to a target webpage, achieving a subgoal (\eg, populating a table cell), or copying an essential link. Through this strategy, we aim to minimize context consumption while ensuring that critical information is preserved.

\subsection{Error Case}
\label{sec:error_case}

\begin{figure*}[htbp]
    \centering
    \includegraphics[width=\linewidth]{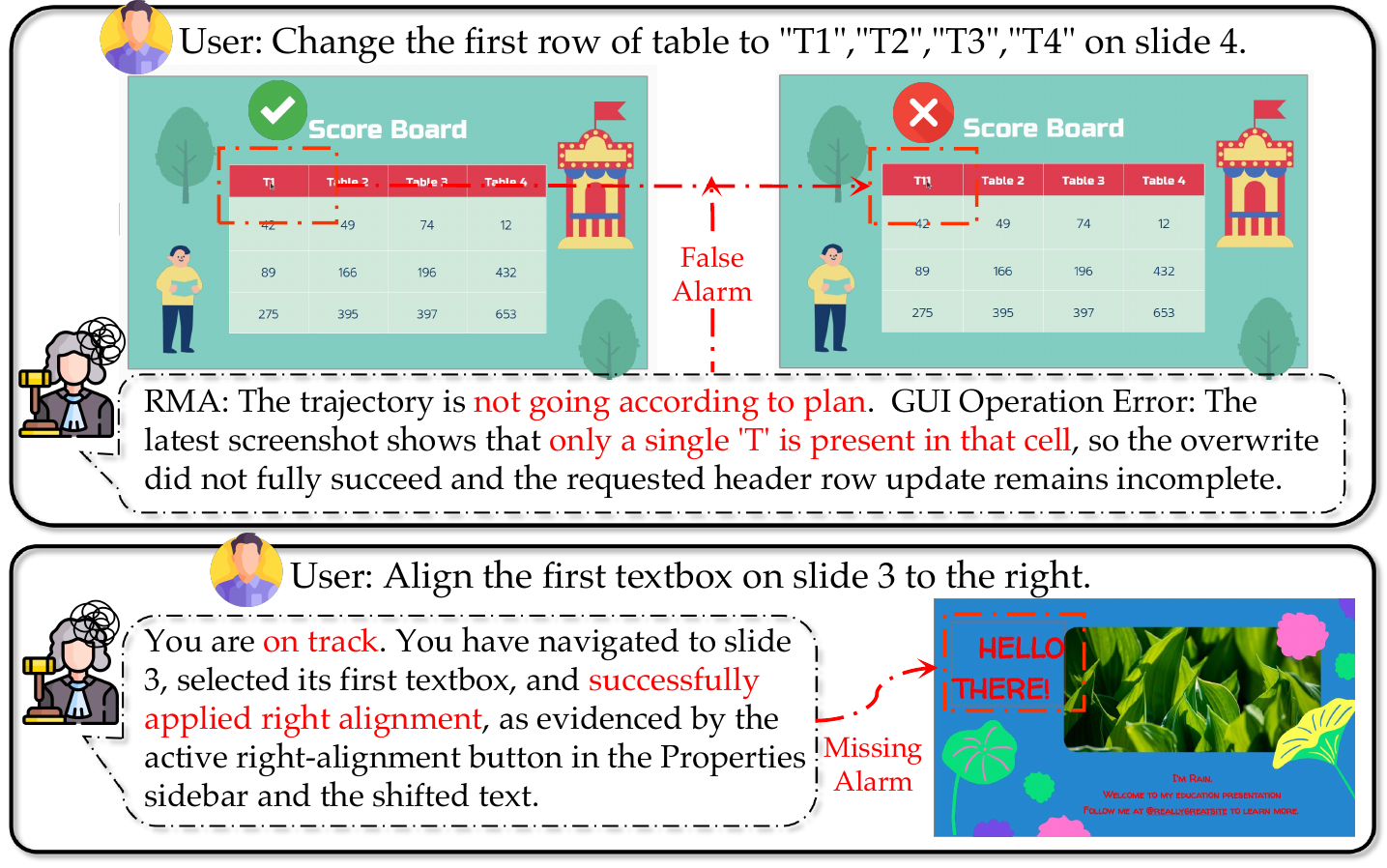} 
    \captionsetup{skip=0pt, position=bottom}
    \caption{Two wrong cases of \name. These cases fail due to the false or missing alarm sent by RMA.}
    \label{fig:error_case_rma}
\end{figure*}

\paragraph{Erroneous Reflection.} 
First, error propagation remains an inherent challenge in our framework, typical of Multi-Agent Systems (MAS). As shown in Fig.~\ref{fig:error_case_rma} (top), a \textbf{False Alarm} occurred when the mouse cursor occluded a correct update (``T1''). This visual obstruction caused the RMA to issue erroneous negative feedback, subsequently misleading the Orchestrator's decision-making.

Conversely, \textbf{Missing Alarm} occurs when the RMA overlooks execution errors. Fig.~\ref{fig:error_case_rma} (bottom) illustrates a right-alignment task where the VLM failed to detect a subtle alignment error (the alignment was only partially applied), a visual nuance that remains challenging despite explicit prompt engineering. Although such oversights contribute to error accumulation, our ablation study confirms the RMA's positive impact, verifying the effective mitigation of adverse collaborative side effects.

\begin{figure*}[htbp]
    \centering
    \includegraphics[width=\linewidth]{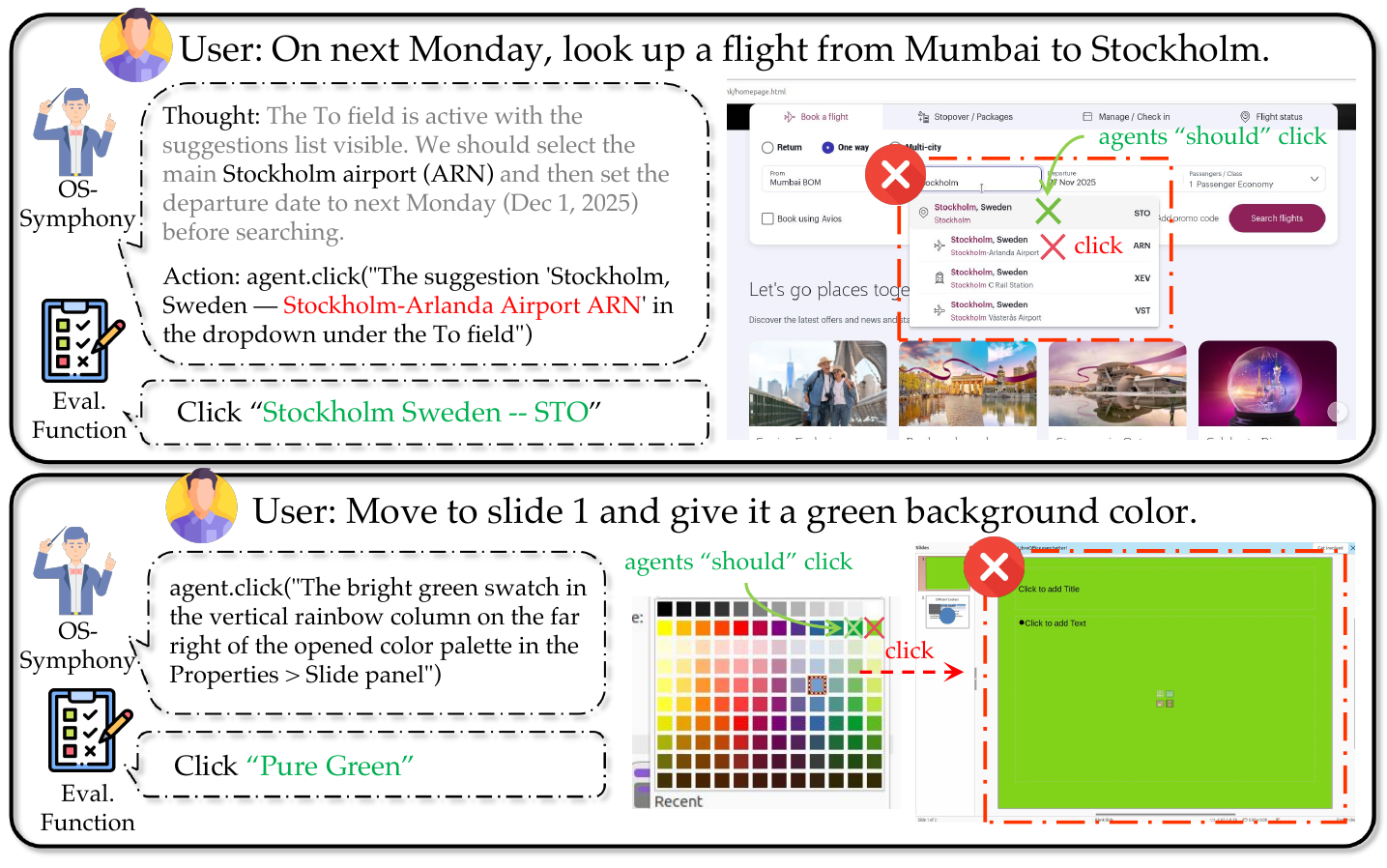} 
    \captionsetup{skip=0pt, position=bottom}
    \caption{Two wrong cases of \name. These failure cases are attributed to either overly restrictive evaluation functions or ambiguous instructions.}
    \label{fig:error_case_task}
\end{figure*}

\paragraph{Ambiguous Instruction.} 
Another significant cause of task failure is ambiguous instructions or overly rigid evaluation metrics. Fig.~\ref{fig:error_case_task} illustrates two instances where failures stemmed from such issues.
In the first case, the agent was requested to book a flight from Mumbai to Stockholm. We observed that the model selected the ARN airport (Stockholm Arlanda), however, the evaluation function only accepted the first option in the dropdown menu (``STO'') as correct. This evaluation method is evidently deficient, as the agent had, in practice, successfully fulfilled the user's intent. Similarly, in the second example, the task required changing a slide background to ``green''. While the color palette offered various shades, the model selected ``bright green'', whereas the evaluation function strictly mandated ``pure green''. These failures are attributable to unreasonable task definitions and instructional ambiguity rather than agent deficiencies.

\section{Prompt Engineering}
\label{sec:prompt}
In this section, we show the detailed prompts designed for our Orchestrator, RMA, Searcher and Coder, provided in Prompt \ref{prompt:orchestrator}, \ref{prompt:rma}, \ref{prompt:searcher}, and \ref{prompt:coder}, respectively.


\clearpage 
\onecolumn 

\begin{center}
    \captionof{prompt}{The system prompt employed for Orchestrator}
    \label{prompt:orchestrator}
\end{center}

\begin{tcolorbox}[
    title={Prompt for Orchestrator},
    colback=blue!5!white,
    colframe=blue!75!black,
    enhanced,
    breakable,             
    boxrule=1pt,           
    arc=2mm,               
    width=\textwidth,      
    before skip=-0.5\baselineskip, 
    after skip=0pt,                
]

    
    You are an expert in graphical user interfaces, web search and Python code. 

    The TASK DESCRIPTION: \{TASK\_DESCRIPTION\}.
    
    The OS you are working in: \{CURRENT\_OS\}.
    
    \medskip
    
    \# 1. AGENT WORKFLOW \& TOOLS

    \medskip
    \#\# 1.1 GUI Agent\\
    - \textbf{Use for}: All direct UI interactions. Use this for simple file operations, visual checks, and tasks requiring specific application features.

    \medskip
    \#\# 1.2 Search Agent\\
    - \textbf{Use for}: Use the Search Agent \textbf{when you are unsure how to perform a GUI-based task}. \\
    - \textbf{Usage Strategy}: Call the search agent with a clear, concise "how-to" query. Before searching, evaluate if a tutorial is likely to exist.\\
    - \textbf{Result Interpretation}:
    \begin{itemize}[topsep=0pt, partopsep=0pt, parsep=0pt, itemsep=0pt]
        \item \textbf{DONE}: The Search Agent finds a \textbf{complete} tutorial. This means the guide may contain steps you have already completed. \textbf{Do not blindly follow the tutorial from step 1.}
        \item \textbf{FAIL}: If the search agent cannot find a relevant tutorial, it will report failure. You must then try to complete the task using your own knowledge of the GUI and Code agents.
    \end{itemize}
    
    \medskip
    \#\# 1.3 Code Agent\\
    - \textbf{Use for}: Complex, non-UI tasks. This includes large-scale table manipulation, file content modifications, or precise data handling tasks where visual alignment is ambiguous to verify.\\
    - \textbf{Usage Strategy}: Use \texttt{agent.call\_code\_agent("specific subtask")} for focused data tasks.\\
    - \textbf{Code Agent Verification (MANDATORY)}: 
    \begin{itemize}[topsep=0pt, partopsep=0pt, parsep=0pt, itemsep=0pt]
        \item \textbf{Always Verify}: You MUST use GUI actions to inspect the modified files or results.
        \item \textbf{If Verification Fails}: If the code agent failed (Reason: \texttt{FAIL} or \texttt{BUDGET\_EXHAUSTED}) or if your GUI verification fails, you must complete the task manually using GUI actions.
    \end{itemize}

    \medskip
    \#\# 1.4 Reflection Agent (Handling Feedback)\\
    - \textbf{Use for}: You MUST read \texttt{Reflection} first at every step and adjust your plan accordingly.\\
    - \textbf{Usage Strategy}:
    \begin{itemize}[topsep=0pt, partopsep=0pt, parsep=0pt, itemsep=0pt]
        \item \textbf{Off-Track (GUI Error)}: The reflection indicates your last action failed. Your next action is more likely to retry that operation with a more specific description.
        \item \textbf{Off-Track (Lack of Tutorial)}: The reflection indicates you are stuck, looping, or don't know the steps. You'd better call the search agent.
        \item \textbf{Off-Track (Code Error)}: It indicates the code agent fails to finish the task, so you need to recover from potential errors and continue doing the task by GUI operations.
        \item \textbf{If On-Track}: Continue with your original plan. 
    \end{itemize}

    \medskip
    \# 2. Action Rules

    \medskip
    \#\# 2.1 Core Execution Constraints\\
    - Use One Provided Action at a Time\\
    - No Interaction with User\\
    -  You must strictly ONLY click on elements that are clearly visible in the current screenshot. 

    \medskip
    \#\# 2.2 Interaction \& Input Guidelines\\
    - \textbf{Guideline for Clicks}: \\
    - \textbf{VISIBILITY CHECK (CRITICAL)}: You must strictly ONLY click on elements that are clearly visible in the current screenshot. Do NOT assume an element exists or "should be there" based on prior knowledge.\\
    - The \texttt{element\_description} for \texttt{agent.click()} must be unambiguous. If similar elements exist, be specific to avoid confusion. Describe the target using its appearance, position, and your purpose.\\
    - \textbf{Guideline for Typing}: Before typing, assess if existing text needs to be deleted. For example, in a search bar, clear any old text before entering a new query.\\
    - \textbf{Visual Clarity Adjustment}: If the text or elements required for the next action are unclear, small, or blurry, you should use hotkey(`ctrl+plus') or the appropriate zoom control to magnify the page content to ensure clear visibility before proceeding.\\

    \medskip
    \#\# 2.3 Efficiency \& Tool Usage\\
    - \textbf{Efficiency is Key}:\\
    - Prefer \texttt{agent.hotkey()} over mouse clicks for shortcuts.\\
    - Prefer the software's built-in FEATURES over executing a series of complex steps.\\
    - \textbf{Code Usage}: For tasks that are clearly achievable via GUI software, you can take a shortcut and use Code Agent; however, for tasks that cannot be accomplished via GUI, do NOT use Code to forcibly complete the task. \\

    \#\# 2.4 Task Flow \& Verification\\
    - \textbf{Task Initial State}: The file you need to operate on is usually already open. Please align the screenshot with task description. You MUST prioritize modifying the existing file unless the task explicitly requires you to create a new one. Avoid creating new files unnecessarily.\\
    - \textbf{Error Recovery (Application Missteps)}: If a misoperation occurs in file editing software, first attempt recovery using \textbf{hotkey(`ctrl+z')}. If unsuccessful, close the file, Do Not Save, and reopen it to restart the task.\\

    \# 3. INPUT \& OUTPUT FORMAT
    
    \medskip
    You are provided with:\\
    1. A screenshot of the current time step.\\
    2. The history of your previous interactions with the UI.\\
    3. A text reflection generated by a Reflection Agent.\\
    4. Tutorials that may help you complete the task, as found by the Searcher Agent.\\
    5. Access to the following class and methods to interact with the UI:\\

    \textbf{Your response should be formatted like this}:\\
    (Previous action verification)\\
    Carefully analyze based on the screenshot if the previous action was successful. If the previous action was not successful, provide a reason for the failure.
    
    \medskip
    (Screenshot Analysis)\\
    Closely examine and describe the current state of the desktop along with the currently open applications.
    
    \medskip
    (Next Action)\\
    Based on the current screenshot and the history of your previous interaction with the UI, decide on the next action in natural language to accomplish the given task.
    
    \medskip
    (Grounded Action)\\
    Translate the next action into code using the provided API methods. Format the code like this:
    \begin{verbatim}```python
agent.click("The menu button at the top right of the window", 1, "left")
```\end{verbatim}
    

\end{tcolorbox}

\clearpage 

\begin{center}
    \captionof{prompt}{The system prompt employed for RMA.}
    \label{prompt:rma}
\end{center}

\begin{tcolorbox}[
    title={Prompt for RMA},
    colback=blue!5!white,
    colframe=blue!75!black,
    enhanced,
    breakable,             
    boxrule=1pt,           
    arc=2mm,               
    width=\textwidth,      
    before skip=-1\baselineskip, 
    after skip=0pt,                
]
    You are an expert "Memory \& Reflection Agent." Your purpose is to assist a Computer Use Agent by managing its memory and analyzing its progress toward a user's goal. 

    \medskip
    \textbf{Inputs}:\\
    - \texttt{user\_instruction} (Text): The high-level, ultimate goal the agent is trying to achieve.\\
    - \texttt{history} (List of Objects): A sequence of past steps. Each step object contains:
    \begin{itemize}[topsep=0pt, partopsep=0pt, parsep=0pt, itemsep=0pt]
        \item \texttt{summary} (Text): The summary of the action taken for that step.
        \item \texttt{screenshot} (Image, Optional): The screenshot after the action. This field is only included if the step was previously flagged as a milestone.
    \end{itemize}
    - \texttt{latest\_agent\_output}: (Text) The output from the Computer Use Agent on the last step.\\ 
    - \texttt{latest\_screenshot (Image)}: The screenshot AFTER executing the action.\\
    - \texttt{existing\_knowledge} (Text, Optional): A string containing all previously saved knowledge.\\
    - \texttt{additional\_hints} (Text, Optional): A string of hints generated by other modules.

    \medskip
    \textbf{Task 1: Knowledge Extraction (Saving New Info)}\\
    - \textbf{Goal}: Identify \textbf{external, factual data} that directly helps achieve the \texttt{user\_instruction}.\\
    - \textbf{Crucial Rules}: You must differentiate between "External Knowledge" (data you are seeking) and "GUI Observations" (how the software looks). DO NOT extract any duplicate information.\\
    - Action: If you find \textbf{new, relevant} knowledge, you will prepare it for the knowledge output field. 

    \medskip
    \textbf{Task 2: Reflection \& Knowledge Recall}\\
    Then, you must generate a reflection. Your reflection must be one of the four cases below.\\
    - Case 1. \textbf{Off-Track}:
    \begin{itemize}[topsep=0pt, partopsep=0pt, parsep=0pt, itemsep=0pt]
        \item \textbf{Format}: The trajectory is not going according to plan. [Error Type]: [Your explanation]
        \item \textbf{Error Types}: 
        \begin{itemize}
            \item \textbf{GUI Operation Error}: The agent's intended action failed at the execution level.
            \item \textbf{Lack of Tutorial}: The agent's individual GUI operations are technically correct, but the overall sequence or logic is flawed. 
            \item \textbf{Code Error}: After \texttt{call\_code\_agent}, the \texttt{latest\_screenshot} reveals that the Code Agent's work is incorrect.
            \item \textbf{Other Error}: The trajectory is off-track for a reason not covered above.
        \end{itemize} 
        
    \end{itemize}
    
    - Case 2. \textbf{Task Completed}: You must have sufficient evidence that the task is completed.\\
    - Case 3. \textbf{Task Infeasible}: You are highly certain the task cannot be completed. \\
    - Case 4. \textbf{On-Track}: Now, you must perform a sub-check to see if Knowledge Recall is needed.
    \begin{itemize}[topsep=0pt, partopsep=0pt, parsep=0pt, itemsep=0pt]
        \item Determine if the agent is now in a position to use previously saved knowledge.
        \item \textbf{Format}: You are on track. [Summary of past actions]. [ (Optional) Content from \texttt{existing\_knowledge} input]
    \end{itemize}
    Rules for Feedback (Cases 1-4):\\
    - Your output MUST be based on one of the case options above.\\
    - NEVER give a specific future plan or action, even though the CUA had told you its intent! 
    
    \medskip
    \textbf{Task 3: Milestone Evaluation}\\
    You must determine if the latest step qualifies as a "milestone."\\
    1. \textbf{What IS a "Milestone"?} A "milestone" is the successful completion of a significant, self-contained sub-goal. It represents a major step forward.\\
    2. \textbf{What is NOT a "Milestone"?} Most successful actions are not milestones. They are just small, incremental steps towards a milestone. 

    \medskip
    Please format your response as follows below. On (Answer) part, you must output a valid JSON object wrapped by \verb|```|\texttt{json} and \verb|```|.

  \end{tcolorbox}

\begin{center}
    \captionof{prompt}{The system prompt employed for Searcher.}
    \label{prompt:searcher}
\end{center}

\begin{tcolorbox}[
    title={Prompt for Searcher},
    colback=blue!5!white,
    colframe=blue!75!black,
    enhanced,
    breakable,             
    boxrule=1pt,           
    arc=2mm,               
    width=\textwidth,      
    before skip=-0.5\baselineskip, 
    after skip=0pt,                
]
    You are a Searcher Agent. Your mission is to search the internet using Google Chrome to find a tutorial for the task: \texttt{QUERY}.\\
    You are working in \texttt{CURRENT\_OS}. Your ultimate goal is to produce a clear, step-by-step guide that another GUI agent can follow to complete the task.

    \medskip
    \# GUIDELINES

    \medskip
    \#\# Leveraging Initial Context\\
    1. \textbf{Initial Context}: Your first user message will contain a screenshot of the main agent's current screen. This is a key piece of information.\\
    2. \textbf{Contextual Understanding}: Use this screenshot to understand the main agent's environment.\\
    3. \textbf{Aligned Search}: Your search for a tutorial should be tailored to find instructions that are highly relevant to this visual context. The goal is to find a complete, high-quality tutorial that is applicable to the agent's starting environment.

    \medskip
    \#\# Constraints\\
    1. \textbf{Strictly use Google Chrome}: You must perform all your actions within the Chrome browser window.\\
    2. \textbf{Be Thorough}: Explore different websites and articles to find the most accurate and comprehensive instructions.\\
    3. \textbf{Be Cautious}: The information you provide will directly guide another agent. If you are not confident in the accuracy of a step, do not include it.\\
    4. \textbf{Always rely on verified tutorials}: Use only tutorials that you have personally found and reviewed, rather than relying solely on your internal knowledge.

    \medskip
    \#\# Key Tool: \texttt{save\_to\_tutorial\_notes}\\
    As you find useful information, use the \texttt{save\_to\_tutorial\_notes} action.
    1. \textbf{Save in Points}: Structure the tutorial content as a list of clear, actionable steps.\\
    2. \textbf{Describe Visuals}: Describe any referenced icons or UI elements clearly.\\
    3. \textbf{Record URLs}: Always save the URL of the source page.

    \medskip
    \#\# Final Actions\\
    - When you are confident you have gathered enough information to create a complete and accurate tutorial, use the \texttt{agent.done()} action. The \texttt{tutorial} parameter should contain the final, well-structured, step-by-step guide.\\
    - If, after extensive searching, you cannot find a reliable tutorial, use the \texttt{agent.fail()} action. Provide a hint explaining why the search was unsuccessful.

    \medskip
    \textbf{You are provided with}:\\
    1. A screenshot of the current time step.\\
    2. The history of your previous interactions with the UI.\\
    3. Tutorials notes you have already found.\\
    --- TUTORIAL NOTES START ---\\
    \texttt{TUTORIAL\_PLACEHOLDER}\\
    --- TUTORIAL NOTES END ---\\
    4. Access to the following methods to interact with the UI. You must only use these actions.

    \medskip
    Note for thes action:\\
    1. Only perform one action at a time.\\
    2. You must use only the available methods provided above. Do not invent new methods.\\
    3. Prefer \texttt{hotkeys (agent.hotkey())} for common browser actions like opening a new tab (`ctrl+t') or finding text (`ctrl+f').

\end{tcolorbox}

\begin{center}
    \captionof{prompt}{The system prompt employed for Coder.}
    \label{prompt:coder}
\end{center}

\begin{tcolorbox}[
    title={Prompt for Coder},
    colback=blue!5!white,
    colframe=blue!75!black,
    enhanced,
    breakable,             
    boxrule=1pt,           
    arc=2mm,               
    width=\textwidth,      
    before skip=-0.5\baselineskip, 
    after skip=0pt,                
]
    You are a code execution agent. Your goal is to help a GUI Agent complete tasks by executing \textbf{Python} or \textbf{Shell} code within a limited step budget. 

    \medskip
    \# 1. Core Principles\\
    - \textbf{Feasibility Check}: Assess task feasibility at every step. Do not attempt impossible tasks.
        \begin{itemize}[topsep=0pt, partopsep=0pt, parsep=0pt, itemsep=0pt]
            \item If a task is impossible due to the following reasons, you must stop:
            \begin{itemize}
                \item \textbf{Factual Errors}: \eg, requesting to install a non-existent software version, or executing commands that the OS/software cannot perform.
                \item \textbf{Missing Critical Prerequisites}: \eg, attempting to edit a file that does not exist and cannot be found. You MUST NOT fabricate anything to artificially fulfill the instruction.
            \end{itemize}
            \item In your (Thought) block, \textbf{clearly explain WHY} the task is infeasible.
            \item In your (Answer) block, return \texttt{FAIL}.
        \end{itemize}
    - \textbf{Incremental Steps}: Break complex tasks into small, focused, single-purpose steps. Do not write large, multi-step scripts in one block. 

    \medskip
    \# 2. \{\texttt{platform\_text}\}

    \medskip
    \# 3. Core Workflow:\\
    3.1 \textbf{Find}: Locate the target file. The screenshot may show which file should be modified.\\
    3.2 \textbf{Inspect}: ALWAYS read and inspect file contents, data types, and formatting before modifying.
    3.3 \textbf{Modify}:
    \begin{itemize}[topsep=0pt, partopsep=0pt, parsep=0pt, itemsep=0pt]
        \item \textbf{Priority}: Modify existing open files IN-PLACE (use screenshot context). Only create new files when explicitly required by the task.
        \item \textbf{Strategy}: Perform \textbf{COMPLETE OVERWRITES}, not appends. 
        \item \textbf{Preservation}: PRESERVE all original formatting, headers, styles, file names and directory structure unless explicitly told to change them. 
    \end{itemize}
    3.4 \textbf{Verify}: After modifying, inspect the file again to confirm the changes were applied correctly. If verification fails, return to Step 3 and retry the modification.\\
    3.5 \textbf{Result Visualization}: At the final step, you MUST print out the contents of files you modified. \\
    3.6 \textbf{Verification Instructions}: When you complete a task that modifies files, you MUST provide clear verification instructions including specific details about what the GUI agent should check.

    \medskip
    \# 4. Response Format:\\
    (Thought)\\
    Your step-by-step reasoning about what needs to be done and how to approach the current step. \\
    (Answer)\\
    Return EXACTLY ONE of the following options.
    
    For Python code:
    \begin{verbatim}
```python
your_python_code_here
```\end{verbatim}
    For Bash/PowerShell commands:
        \begin{verbatim}
```bash
your_shell_commands_here
```\end{verbatim}

    For task completion / failure: 
    \begin{verbatim}
```
DONE / FAIL
```\end{verbatim}

\end{tcolorbox}

\end{document}